\documentclass[11pt,a4paper]{article}
\usepackage[a4paper,margin=22mm]{geometry}
\usepackage{graphicx}
\usepackage{multirow}
\usepackage{amsmath,amssymb,amsfonts}
\usepackage{amsthm}
\usepackage{mathrsfs}
\usepackage[title]{appendix}
\usepackage{xcolor}
\usepackage{textcomp}
\usepackage{manyfoot}
\usepackage{booktabs}
\usepackage{algorithm}
\usepackage{algorithmicx}
\usepackage{algpseudocode}
\usepackage{listings}
\usepackage{float}
\usepackage{tabularx,colortbl}
\usepackage{siunitx}
\usepackage{caption}
\usepackage{subcaption}
\usepackage[hidelinks]{hyperref}
\usepackage{placeins}
\usepackage{flafter}
\usepackage{needspace}
\usepackage{microtype}
\usepackage{enumitem}
\graphicspath{{Figures/}}
\newlength{\bibsep}

\begin{document}
\hypersetup{
pdftitle={Robust inverse dielectric design of skin phantoms for implantable wireless link evaluation},
pdfauthor={Andreas Prokscha, Ali Alhaj Abbas, Samar Elmaadawy, Mandana Jalali, Yamen Zantah, Josep M. Jornet, Daniel Erni, Thomas Kaiser},
pdfsubject={Preprint},
pdfkeywords={implantable wireless communication, tissue phantoms, robust inverse dielectric design, terahertz communication}
}
\begin{center}
{\LARGE\bfseries Robust inverse dielectric design of skin phantoms for implantable wireless link evaluation\par}
\vspace{0.8em}
{\large Andreas Prokscha$^{1,*}$, Ali Alhaj Abbas$^{2,*}$, Samar Elmaadawy$^{1,3}$, Mandana Jalali$^{4}$, Yamen Zantah$^{1}$, Josep M. Jornet$^{3}$, Daniel Erni$^{4}$, and Thomas Kaiser$^{1}$\par}
\vspace{0.65em}
{\small
$^{1}$Institute of Digital Signal Processing (DSV), University of Duisburg-Essen (UDE), Duisburg 47057, Germany\\
$^{2}$Interdisciplinary Research Center for Communication Systems and Sensing, King Fahd University of Petroleum \& Minerals (KFUPM), Dhahran 31261, Saudi Arabia\\
$^{3}$Department of Electrical and Computer Engineering, Northeastern University, Boston, MA 02115, USA\\
$^{4}$General and Theoretical Electrical Engineering (ATE), Faculty of Engineering, University of Duisburg-Essen (UDE) and Center for Nanointegration Duisburg-Essen (CENIDE), Duisburg 47048, Germany\\[0.45em]
$^{*}$Corresponding authors: Andreas Prokscha (andreas.prokscha@uni-due.de) and Ali Alhaj Abbas (ali.abbas@kfupm.edu.sa)
}\par
\vspace{0.7em}
{\small\bfseries Preprint}\par
\end{center}
\vspace{0.5em}
\begin{abstract}
Skin phantoms support reproducible evaluation of implantable wireless links, yet published complex permittivities of skin predict different propagation losses. We formulate a robust inverse design to determine a homogeneous phantom permittivity that minimizes the largest channel-loss difference across selected skin reference models, incidence angles and both transverse-electric and transverse-magnetic polarizations. Computed candidates and analytical lower bounds bracket the optimal mismatch and constrain the critical depth. Across seven frequencies from 5 to 500~GHz, the minimax mismatch at 1.3~mm ranges from 1.13 to 13.72~dB. For a 3~dB tolerance, suitable permittivities exist at every investigated depth at 5~GHz, while critical depths at the other frequencies range from 0.23 to 1.01~mm. Only one of seven material cases from the literature meets a 6~dB tolerance at 1.3~mm. These results translate channel-loss tolerances into quantitative design criteria for skin phantoms used in implantable wireless link evaluation.
\end{abstract}
\noindent\textbf{Keywords:} Body-centric propagation, implantable wireless communication, inverse dielectric design, robust optimization, terahertz communication, tissue phantoms
\vspace{0.8em}

\section{Introduction}
\label{sec:introduction}

\begin{figure}[!htbp]
\centering
\includegraphics[width=\textwidth]{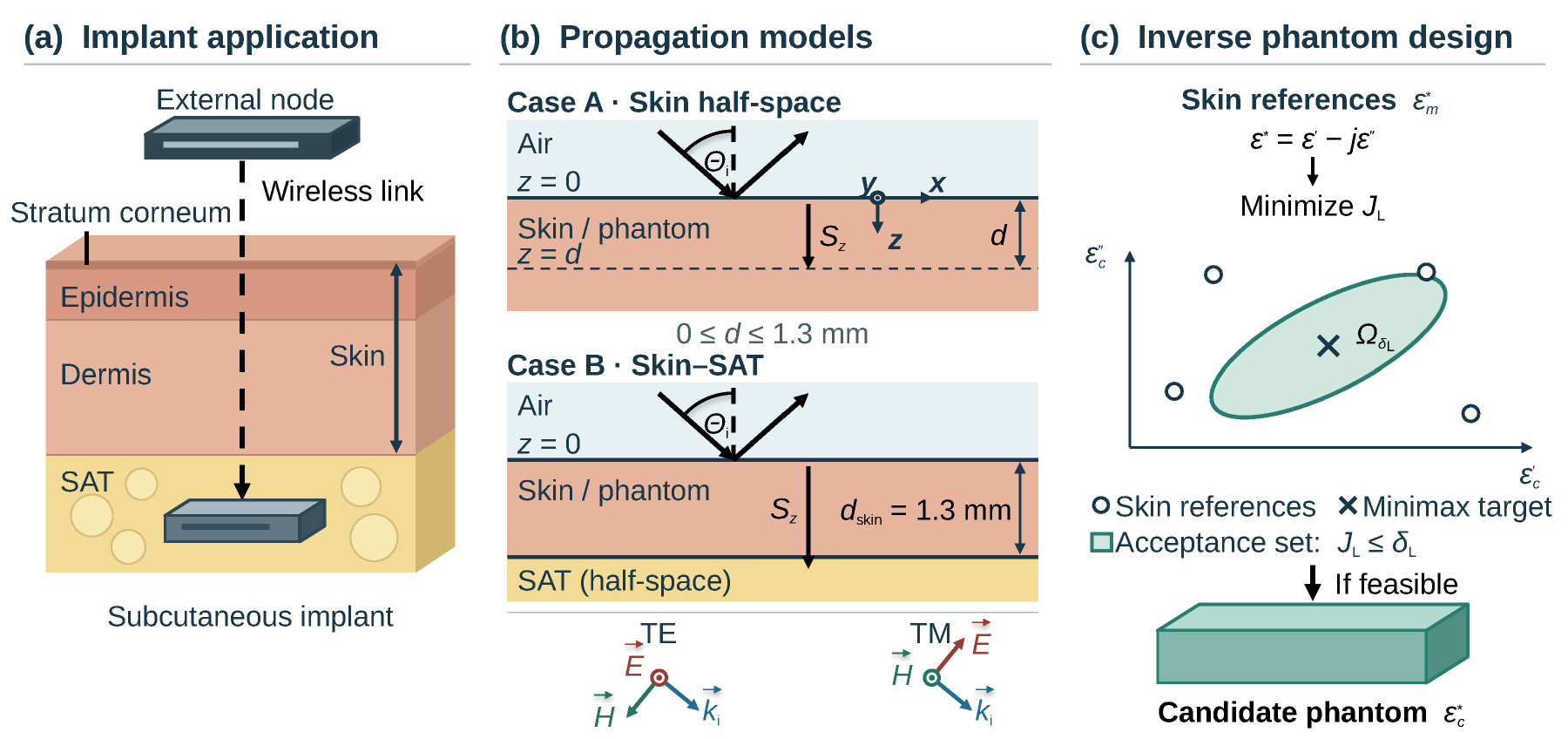}
\caption{Application context, propagation models and inverse phantom design. \textbf{(a)} Wireless link between an external node and a subcutaneous implant. The stratum corneum is part of the epidermis. \textbf{(b)} Identical planar models for skin and phantom. Case A: homogeneous half-space with observation plane $z=d$, $0\leq d\leq1.3~\mathrm{mm}$. Case B: skin layer of thickness $d_{\mathrm{skin}}=1.3~\mathrm{mm}$ over semi-infinite SAT with fixed permittivity for each SAT model. Power transfer into SAT is evaluated at the lower interface. The incident wavevector $\vec{k}_{\mathrm{i}}$ lies in the $x$--$z$ plane at angle $\Theta_{\mathrm{i}}$ to the surface normal, with positive $y$ pointing out of the drawing. For TE and TM polarization, the incident electric and magnetic field vectors, respectively, are parallel to the unit vector $\hat{e}_y$. The symbol $S_z$ denotes the normal time-averaged power flux density. \textbf{(c)} Illustrative acceptance set in the complex permittivity plane. Circles represent skin references, the cross marks a minimax target, and shading denotes the set $\Omega_{\delta_{\mathrm{L}}}$ satisfying $J_{\mathrm{L}}\leq\delta_{\mathrm{L}}$.}
\label{fig:framework}
\end{figure}

Implantable wireless-link experiments depend on controlled representations of the electromagnetic (EM) response of biological tissue. Characterized skin phantoms provide repeatable dielectric conditions and complement computational tissue models~\cite{Joseph2025}. For wireless communication between an external device and an implant located within the subcutaneous adipose tissue (SAT), as illustrated in Fig.~\ref{fig:framework}(a), the overlying skin constitutes an essential part of the propagation channel. Reflection at the air--skin interface and absorption within skin reduce the transmitted power. This contribution to the link budget is modeled using plane-wave incidence from air, with power transfer evaluated at a normal depth $d$ below the skin surface. Under otherwise identical link conditions, an increase in propagation loss reduces the received signal-to-noise ratio (SNR). Figure~\ref{fig:framework} connects this application to the two planar models used for skin and phantom. Case A (skin half-space) describes propagation in a homogeneous medium over observation depths of $0\leq d\leq1.3~\mathrm{mm}$. The upper limit of this depth range is motivated by ultrasound measurements of adult forearm skin thickness~\cite{VanMulder2017}. Case B (skin--SAT model) describes a 1.3~mm skin layer over semi-infinite SAT. Power transfer into SAT is evaluated at the lower interface, with its permittivity fixed across all references and phantom candidates for each investigated SAT model. Case A forms the basis of the inverse design, while Case B examines sensitivity to the underlying tissue.

For such propagation studies, previous research has established broadband, reflection-oriented and multilayer phantom materials for body-centric propagation~\cite{Chahat2012,GuraliucTMTT2014,GuraliucMobiHealth2014,Khorshid2018,Khorshid2019}. Characterization experiments at Ka- and D-band frequencies and scattering measurements at steep millimeter-wave (mmWave) incidence angles provide further insight into artificial human tissues and skin phantoms~\cite{ProkschaIWMTS2023,ZauberIRMMW2025}. Equivalent homogeneous tissue properties have been examined for implantable-antenna testing and estimated using transmission-matrix and finite-element methods~\cite{GhoshGupta2026,GhoshEtAl2026}. Recent terahertz (THz) phantom materials expand the experimentally characterized options~\cite{Mizuno2024,Yamazaki2025}. Beyond biomedical applications, dielectric characterization combined with scattering analysis has supported the assessment of biological surrogates for THz sensing of insects~\cite{Prokscha2025HoneyBee}. The intra-body technology roadmap emphasizes benchmark phantoms and shared validation data~\cite{Skos2026}, while IEEE~802.15.6-2026 provides the context of wireless body-area networks~\cite{IEEE8021562026}. Agreement at the skin surface alone cannot establish loss agreement at finite depth. Phantom design must therefore account for the propagation requirements of the intended communication experiment~\cite{Pellegrini2013,Yang2015,Abbasi2016}. Published dielectric functions of skin differ with anatomical site, specimen condition, measurement method and frequency range~\cite{Sasaki2022,Wang2021,Wu2015}. In vivo measurements of porcine abdominal and thoracic tissues at microwave frequencies further document tissue variation and measurement uncertainty~\cite{FornesLeal2019}. Stochastic analyses of implant radiation efficiency and numerical studies of communication link margins demonstrate the relevance of dielectric uncertainty to implantable systems~\cite{Sipus2021,Ghosh2025Link}. A phantom specified by a single published reference may consequently reproduce a different propagation loss from that predicted by other references. This connection raises two fundamental questions for phantom design: can a prescribed channel-loss tolerance be met across the selected skin reference models, and, if so, which complex permittivity of a homogeneous phantom can achieve it? The answers depend on transmission at the air--skin interface, attenuation within skin, incidence angle, polarization and observation depth. We develop an analytical and numerical framework for robust inverse phantom design that combines electromagnetic propagation models, minimax optimization and analytical bounds on achievable channel-loss fidelity. At each carrier frequency $f$, a separate reference set $\mathcal{M}(f)$ is defined. Each reference model $m$ is characterized by its complex relative electrical permittivity (i.e. the dielectric function) $\varepsilon_m^*(f)$. For each phantom candidate, the complex permittivity is given by $\varepsilon_{\mathrm{c}}^*=\varepsilon_{\mathrm{c}}'-j\varepsilon_{\mathrm{c}}''$, where $\varepsilon_{\mathrm{c}}'$ and $\varepsilon_{\mathrm{c}}''$ denote the real and nonnegative loss components, respectively. A candidate selected as a material specification is referred to as a target. Within the homogeneous design class $\mathcal{D}(f)$, the permittivities satisfy $\varepsilon_{\mathrm{c}}'>0$ and $\varepsilon_{\mathrm{c}}''\geq0$. Both transverse-electric (TE, $s$-polarization) and transverse-magnetic (TM, $p$-polarization) waves are considered in the primary design. The inverse problem in Fig.~\ref{fig:framework}(c) minimizes the largest absolute loss difference $J_{\mathrm{L}}(\varepsilon_{\mathrm{c}}^*;f,d)$ between the candidate and the references across the prescribed incidence angles and polarizations at frequency $f$ and depth $d$. Its infimum over the design class is $J_{\mathrm{L}}^\star(f,d)$. For a specified loss tolerance $\delta_{\mathrm{L}}$, the acceptance set $\Omega_{\delta_{\mathrm{L}}}(f,d)$ contains all candidate permittivities satisfying $J_{\mathrm{L}}\leq\delta_{\mathrm{L}}$. This provides a material specification that extends beyond a single optimum. An empty acceptance set signifies that the prescribed tolerance is unattainable within the design class. The critical depth $d_{\mathrm{crit}}$ is the infimum of depths for which this occurs. Together, these criteria distinguish material-specific deviations from feasibility limits imposed by the selected skin reference models and the homogeneous design class (cf. Section~\ref{subsec:feasibility}). A guide to notation and terminology is provided in the Supplementary Information. We evaluate seven carrier frequencies independently (5, 28, 60, 77, 110, 300 and 500~GHz), with incidence angles from $0^\circ$ to $85^\circ$ in $1^\circ$ steps and one common target for both polarizations at each frequency and depth. The illustrative tolerances of 1, 3 and 6~dB quantify the influence of the permitted loss mismatch on material selection and feasibility. Comparison with published materials illustrates how these criteria guide material selection. The results establish frequency- and depth-dependent feasibility limits for homogeneous skin phantoms, with numerical solutions closely approaching the analytical lower bounds across all seven carrier frequencies. Comparisons with published phantom materials and sensitivity analyses of reference selection, incidence angles and tissue boundaries illustrate how these limits guide material selection for implantable wireless link evaluation.

\section{Results}
\label{sec:results}

\subsection{Achievable channel-loss fidelity depends on the skin reference models}
\label{subsec:finite_depth_results}

Before selecting a phantom material for an implantable wireless link experiment, it is essential to establish which loss tolerances can be satisfied across the selected skin reference models. For an observation depth of 1.3~mm, robust inverse design seeks the complex permittivity that minimizes the largest loss mismatch across all prescribed incidence angles and both polarizations. The resulting joint TE/TM minimax mismatch ranges from 1.13 to 13.72~dB, with the computed values lying within 0.001~dB of the analytical lower bound $J_{\mathrm{LB}}$ at all seven carrier frequencies (Table~\ref{tab:headline} and Supplementary Fig.~S1). This close agreement indicates that, at each frequency, a common candidate permittivity approaches the limit imposed by the reference-loss range across all prescribed incidence angles and both polarizations (cf. Section~\ref{subsec:feasibility}). Each frequency constitutes a distinct inverse problem because $\mathcal{M}(f)$ varies. At 300~GHz, five reference models yield a minimax mismatch of 11.72~dB, compared with 3.86~dB for three models at 500~GHz (cf. Table~\ref{tab:headline}). The corresponding channel-loss ranges at normal incidence and $d=1.3~\mathrm{mm}$ are 38.48--59.45 and 60.48--67.57~dB, respectively (cf. Supplementary Table~S14). Despite higher absolute channel loss at 500~GHz, the reference models predict a narrower loss range. Minimax optimization extends this comparison to all prescribed incidence angles and both polarizations.

\begin{figure}[!htbp]
\centering
\includegraphics[width=0.97\columnwidth]{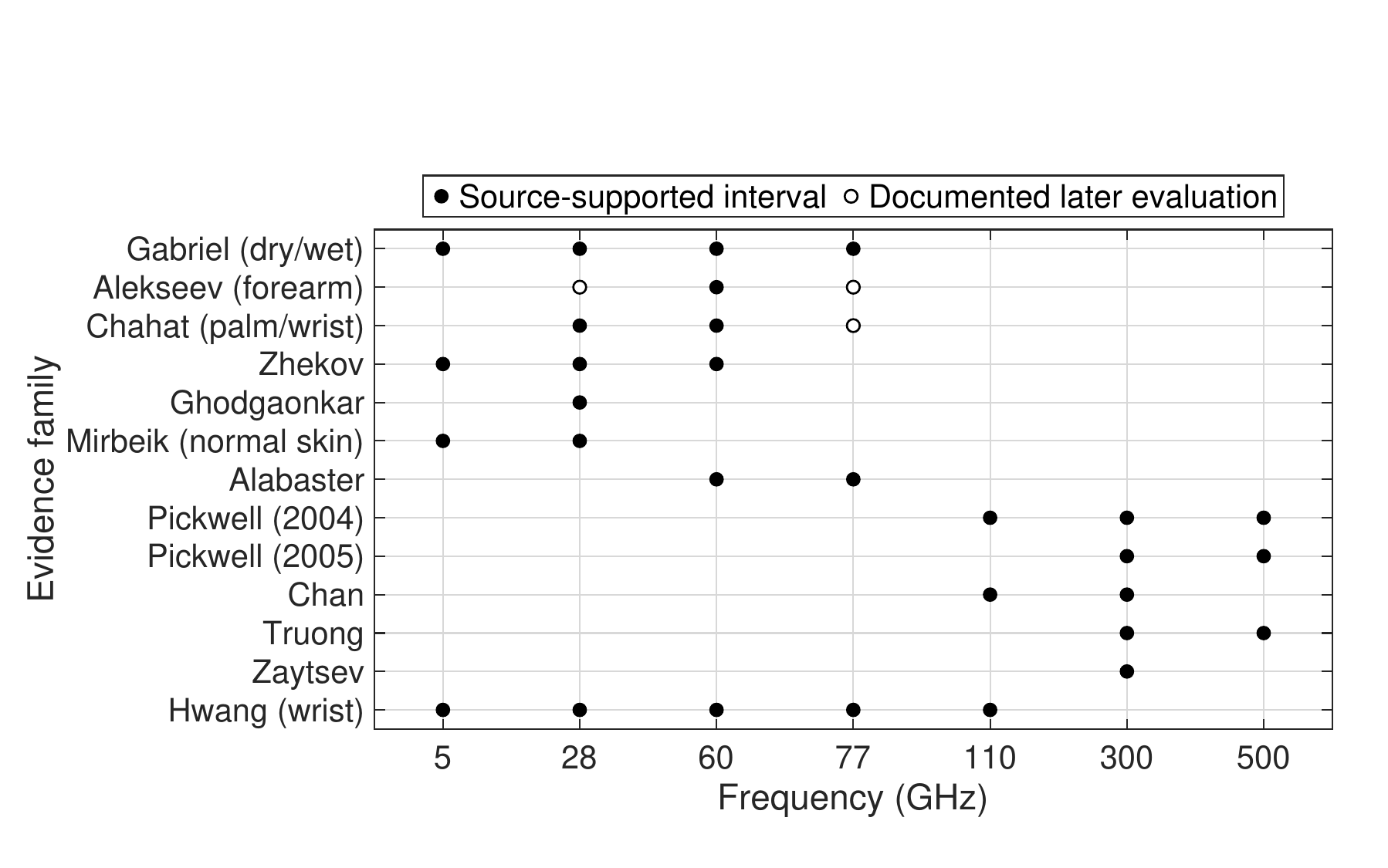}
\caption{Skin reference data at each frequency used in $\mathcal{M}(f)$. Filled markers indicate frequency coverage reported in the original source and open markers indicate a documented later evaluation. Blank cells indicate that a source family does not contribute at that frequency. Rows represent groups of related literature models, each of which can contain multiple reference models.}
\label{fig:evidence}
\end{figure}

\begin{figure}[!t]
\centering
\includegraphics[width=0.95\textwidth]{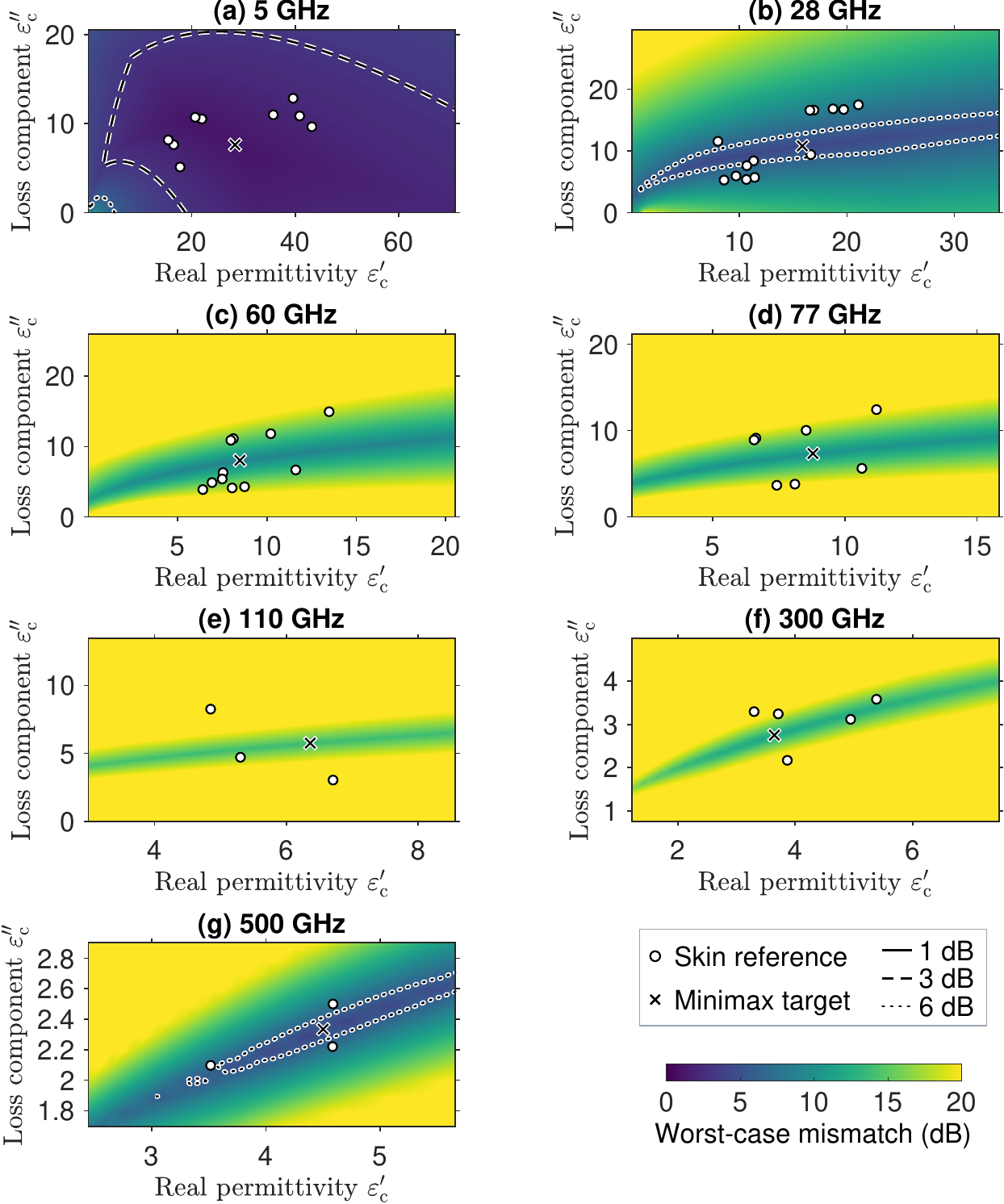}
\caption{Joint TE/TM worst-case channel-loss mismatch maps $J_{\mathrm{L}}(\varepsilon_{\mathrm{c}}',\varepsilon_{\mathrm{c}}'';f,d)$ at $d=1.3~\mathrm{mm}$ for (a) 5, (b) 28, (c) 60, (d) 77, (e) 110, (f) 300, and (g) 500~GHz. Coordinates show the real permittivity and nonnegative loss component of $\varepsilon_{\mathrm{c}}^\ast=\varepsilon_{\mathrm{c}}'-j\varepsilon_{\mathrm{c}}''$. Open circles denote selected skin reference models and crosses denote robust minimax targets. Solid, dashed, and dotted contours represent 1, 3, and 6~dB mismatch wherever these levels occur within the displayed range. Colors indicate interpolated mismatch values, with the color scale limited to 20~dB. Contours are derived from mismatch values calculated on the permittivity grid.}
\label{fig:maps13}
\end{figure}

\begin{table}[!htbp]
\caption{Joint TE/TM robust inverse-design results at $d=1.3~\mathrm{mm}$. Computed mismatch values approximate $J_{\mathrm{L}}^\star$ within the prescribed optimization accuracy. The reported permittivities provide numerically near-optimal design targets, while allowing for multiple material solutions with comparable channel-loss fidelity. Four decimal places are retained for numerical reproducibility. $N_f$ denotes the number of skin reference models at frequency $f$. The final column lists the loss tolerances for which $\Omega_{\delta_{\mathrm{L}}}(f,d)$ is nonempty. A dash indicates that none of the investigated tolerances is feasible.}
\label{tab:headline}
\centering
\scriptsize
\setlength{\tabcolsep}{2.8pt}
\begin{tabular*}{0.88\textwidth}{@{\extracolsep{\fill}}rc|cccc}
\toprule
$f$ &
$N_f$ &
$J_{\mathrm{L}}^\star$ &
$\varepsilon_{\mathrm{c}}'$ &
$\varepsilon_{\mathrm{c}}''$ &
Feasible $\delta_{\mathrm{L}}$ \\
(GHz) & & (dB) & & & (dB) \\
\midrule
5   & 9  & 1.13 & 28.3759 & 7.6279 & 3, 6 \\
28  & 13 & 4.37 & 15.8567 & 10.7977 & 6 \\
60  & 11 & 9.01 & 8.4884 & 8.0509 & -- \\
77  & 7  & 10.21 & 8.7923 & 7.3421 & -- \\
110 & 3  & 13.72 & 6.3618 & 5.7539 & -- \\
300 & 5  & 11.72 & 3.6450 & 2.7515 & -- \\
500 & 3  & 3.86 & 4.5019 & 2.3351 & 6 \\
\bottomrule
\end{tabular*}
\end{table}

For a given incidence angle and polarization, half the difference between the minimum and maximum reference losses provides an analytical lower bound on the achievable mismatch (cf. Section~\ref{subsec:feasibility}). At all seven frequencies at $d=1.3~\mathrm{mm}$, the largest value of this bound is attained within the prescribed $1^\circ$ angular grid for TE polarization at $85^\circ$, although the minimax candidate must also satisfy all other angular and polarization constraints. Figure~\ref{fig:evidence} summarizes the literature sources contributing to each frequency-specific reference set, while Supplementary Table~S6 identifies the reference pairs defining the analytical
lower bounds. At 300~GHz, the Chan and Zaytsev healthy-skin models yield a lower bound of 11.72~dB, which the computed
candidate objective exceeds by only $2.12\times10^{-4}$~dB. The consistently small gaps between the candidate objectives and their analytical lower bounds establish numerical near-optimality across all seven reference sets under the specified evaluation conditions. Filled markers indicate support within the original reported frequency interval and open markers indicate documented later evaluations. Each source family can contain several models, as reflected in the individual model count $N_f$ in Table~\ref{tab:headline}. The largest reference set comprises 13 models at 28~GHz, while three models are included at both 110 and 500~GHz. Supplementary Table~S5 provides the corresponding complex permittivities for all frequencies. Selection rules and records are described in Sections~\ref{subsec:evaluation_framework} and~\ref{sec:biological_evidence}. Figure~\ref{fig:maps13} shows the mismatch as a function of phantom permittivity at 1.3~mm. Contours illustrate the acceptance set within finite plotting windows. Feasibility follows from a candidate meeting the tolerance, while $J_{\mathrm{LB}}>\delta_{\mathrm{L}}$ proves infeasibility independently of the plotted range (Section~\ref{subsec:feasibility}). At an observation depth of 1.3~mm, the minimum achievable joint TE/TM mismatch exceeds 1~dB at all seven frequencies. Suitable permittivities exist at 5~GHz for 3~dB and additionally at 28 and 500~GHz for 6~dB. At 500~GHz, the narrow, tilted region of low mismatch illustrates a trade-off between real permittivity and the loss component. A change in one component may be offset by a change in the other because both interface transmission and attenuation depend on complex permittivity. The acceptance set therefore specifies admissible combinations of both permittivity components. At 60, 77, 110 and 300~GHz, the minimum mismatch exceeds 6~dB, so none of the three tolerance contours occurs. Supplementary Fig.~S2 shows how the mismatch maps differ at the shallower depth of 0.1~mm.

\subsection{Application to published phantom materials}
\label{subsec:published_phantom}

To assess the practical value of the criterion for material selection, seven material cases from six published studies are evaluated (Table~\ref{tab:external_material_screening}). Each fixed material permittivity is evaluated in Case A at 0.1 and 1.3~mm under the original angular and joint TE/TM conditions. The evaluations at 300 and 500~GHz use the same phantom formulation reported by Mizuno et al.~\cite{Mizuno2024}, with the corresponding complex permittivity at each frequency. Table~\ref{tab:external_material_screening} distinguishes permittivity values explicitly reported in the cited sources
(Direct), published fits evaluated within their reported measurement ranges (Fit), values estimated from published graphs (Graph), and published models evaluated beyond their reported measurement ranges (Extrapolation).

\begin{table}[!htbp]
\caption{Comparison of seven material cases from the literature at the selected frequencies under the joint TE/TM mismatch criterion. $J_{0.1}$ and $J_{1.3}$ denote the worst-case mismatch (dB) at 0.1 and 1.3~mm. The last column gives the nominal 6~dB decision at 1.3~mm. Input provenance and full TE/TM results are provided in Supplementary Tables~S10--S12.}
\label{tab:external_material_screening}
\centering
\footnotesize
\setlength{\tabcolsep}{3.5pt}
\renewcommand{\arraystretch}{1.12}
\begin{tabular}{@{}r l c r r c@{}}
\toprule
$f$ (GHz) & Material study & Basis & $J_{0.1}$ & $J_{1.3}$ & 6~dB \\
\midrule
5 & Costanzo \emph{et al.}~\cite{Costanzo2021} & Graph & 1.77 & 2.09 & Pass \\
28 & Mirbeik-Sabzevari \& Tavassolian~\cite{Mirbeik2019} & Fit & 2.92 & 7.92 & Fail \\
60 & Chahat \emph{et al.}~\cite{Chahat2012} & Direct & 2.92 & 16.75 & Fail \\
77 & L\'a\v{c}\'ik \emph{et al.}~\cite{Lacik2016} & Graph & 1.92 & 13.39 & Fail \\
110 & Christ \emph{et al.}~\cite{Christ2025} & Extrapolation & 1.92 & 15.18 & Fail \\
300 & Mizuno \emph{et al.}~\cite{Mizuno2024} & Graph & 2.56 & 25.92 & Fail \\
500 & Mizuno \emph{et al.}~\cite{Mizuno2024} & Graph & 3.51 & 33.64 & Fail \\
\bottomrule
\end{tabular}
\end{table}

For an observation depth of 1.3~mm, the material evaluated at 5~GHz meets the illustrative 6~dB tolerance. Although suitable homogeneous targets exist at 28 and 500~GHz, the selected materials exceed this tolerance. For the remaining frequencies of 60, 77, 110 and 300~GHz, even the minimum achievable mismatch exceeds 6~dB. For example, the Chahat 60~GHz phantom ($\varepsilon_{\mathrm c}^{\ast}=7.4-j11.4$) yields 2.92~dB at 0.1~mm and 16.75~dB at 1.3~mm, compared with the latter depth's optimized 9.01~dB. For this fixed material, the estimated depths at which the mismatch first exceeds the tolerance are 0.11 and 0.37~mm for 3 and 6~dB (Supplementary Table~S9), distinct from the certified limits obtained by optimizing the permittivity separately at each depth. The complex permittivities used for material screening are obtained from reported numerical values at 60~GHz, a published Cole--Cole model within its measurement range at 28~GHz, and estimates from published dielectric or optical-property spectra at 5, 77, 300 and 500~GHz. At 110~GHz, the Cole--Cole model of the validation gel is evaluated beyond its measured permittivity range of 5--67~GHz, with additional context from a separate coated-waveguide reflection comparison (cf. Supplementary Table~S10). The channel-loss mismatch is calculated independently for each material. Sensitivity analysis confirms that the acceptance decisions for a 6~dB tolerance at $d=1.3~\mathrm{mm}$ remain unchanged across the investigated parameter ranges, while the decisions for a 3~dB tolerance at $d=0.1~\mathrm{mm}$ depend on the estimated material parameters at 300 and 500~GHz (cf. Supplementary Table~S12).

\subsection{Feasibility depends on depth and loss tolerance}
\label{subsec:validity_depth}

\begin{figure}[!htbp]
\centering
\includegraphics[width=0.90\columnwidth]{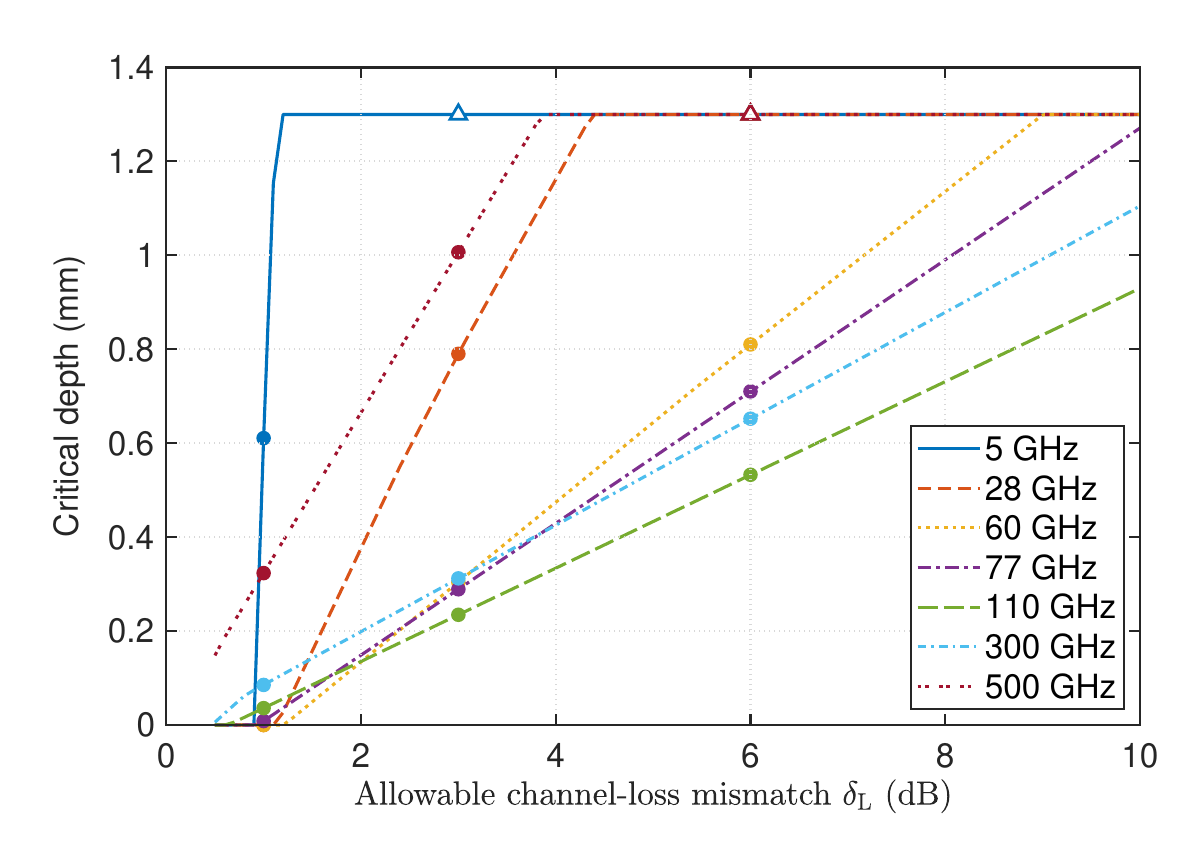}
\caption{Critical depth $d_{\mathrm{crit}}$ versus allowable loss mismatch $\delta_{\mathrm{L}}$. Filled circles mark the certified results at 1, 3 and 6~dB. Open triangles indicate feasibility throughout the modeled interval up to 1.3~mm. Curves visualize the interpolated tolerance dependence. Each depth permits a separately optimized common TE/TM target.}
\label{fig:dcrit_continuous}
\end{figure}

The results at $d=1.3~\mathrm{mm}$ establish which loss tolerances are attainable at the modeled skin thickness. Extending the inverse design to observation depths of $0\leq d\leq1.3~\mathrm{mm}$ reveals how the feasibility limits vary with the prescribed loss tolerance (cf. Fig.~\ref{fig:dcrit_continuous}). At each depth, the target may be optimized independently. A candidate whose mismatch is at or below the tolerance proves that a suitable permittivity exists at that depth. Conversely, an analytical lower bound above the tolerance establishes infeasibility. Combining these tests over depth intervals encloses the critical depth (Section~\ref{subsec:feasibility}).

For a 3~dB tolerance, the critical-depth estimates span 0.23--1.01~mm over the carrier frequencies from 28 to 500~GHz, whereas the 5~GHz case remains feasible throughout the 1.3~mm evaluation interval. Across the 21 frequency--tolerance combinations, the analysis identifies 15 finite intervals containing the critical depth, four cases that meet the tolerance throughout the investigated range, and two that are already infeasible at the surface. Differences in interface transmission impose limits even before propagation attenuation accumulates. The analytical surface lower bounds are 1.14~dB at 28~GHz and 1.21~dB at 60~GHz, making a 1~dB loss tolerance infeasible at both carrier frequencies. This constraint arises from the spread of modeled air--skin interface transmission across the selected references and incidence angles. With increasing depth, differences in attenuation add to the mismatch. Supplementary Fig.~S3 distinguishes median absolute reference skin loss at normal incidence from the jointly optimized worst-case mismatch over depth. The certified critical depths for loss tolerances of 1, 3 and 6~dB are listed in Supplementary Table~S17. Increasing the tolerance from 3 to 6~dB at 28~GHz extends the range over which suitable permittivities exist from approximately 0.79~mm to the entire investigated interval of 1.3~mm. For 110 and 300~GHz, the corresponding critical-depth estimates at a tolerance of 6~dB are 0.53 and 0.65~mm, respectively. Despite the higher absolute skin loss at 500~GHz, the same tolerance can be satisfied throughout the investigated interval. As shown in Section~\ref{subsec:finite_depth_results}, the selected reference models at 500~GHz predict a narrower range of losses at normal incidence than those at 300~GHz. Achievable phantom fidelity depends on the differences between predicted reference losses across the specified propagation conditions.

\subsection{Material selection and fidelity at a given depth}
\label{subsec:benchmark_results}

\begin{figure}[!htbp]
\centering
\includegraphics[width=0.78\columnwidth]{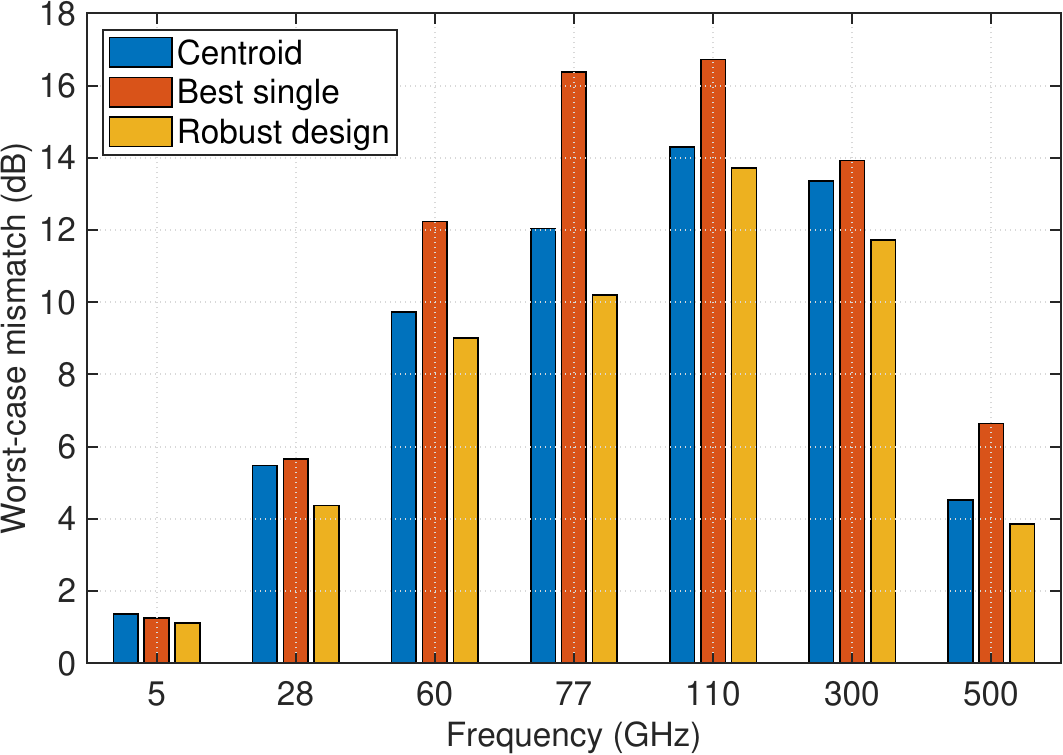}

\caption{Worst-case channel-loss mismatch at $d=1.3~\mathrm{mm}$ for the centroid with equal weighting of provenance groups, best single-reference target, and polarization-specific robust design. Each bar shows the larger of the separately evaluated TE and TM mismatches. The best single-reference and robust targets
may differ between TE and TM.}
\label{fig:strategy_comparison}
\end{figure}

To assess the benefit of robust inverse design for material selection, Fig.~\ref{fig:strategy_comparison} compares three strategies at an observation depth of 1.3~mm. One approach constructs a representative permittivity, referred to as the centroid, by averaging the complex permittivities within each provenance group and assigning equal weight to the resulting group means. Alternatively, an individual skin reference model can be selected by identifying the published permittivity that yields the smallest worst-case mismatch across the entire reference set. Robust inverse design generalizes this approach by minimizing the worst-case channel-loss mismatch over the continuous space of admissible homogeneous permittivities.

\begin{figure}[!htbp]
\centering
\includegraphics[width=0.78\columnwidth]{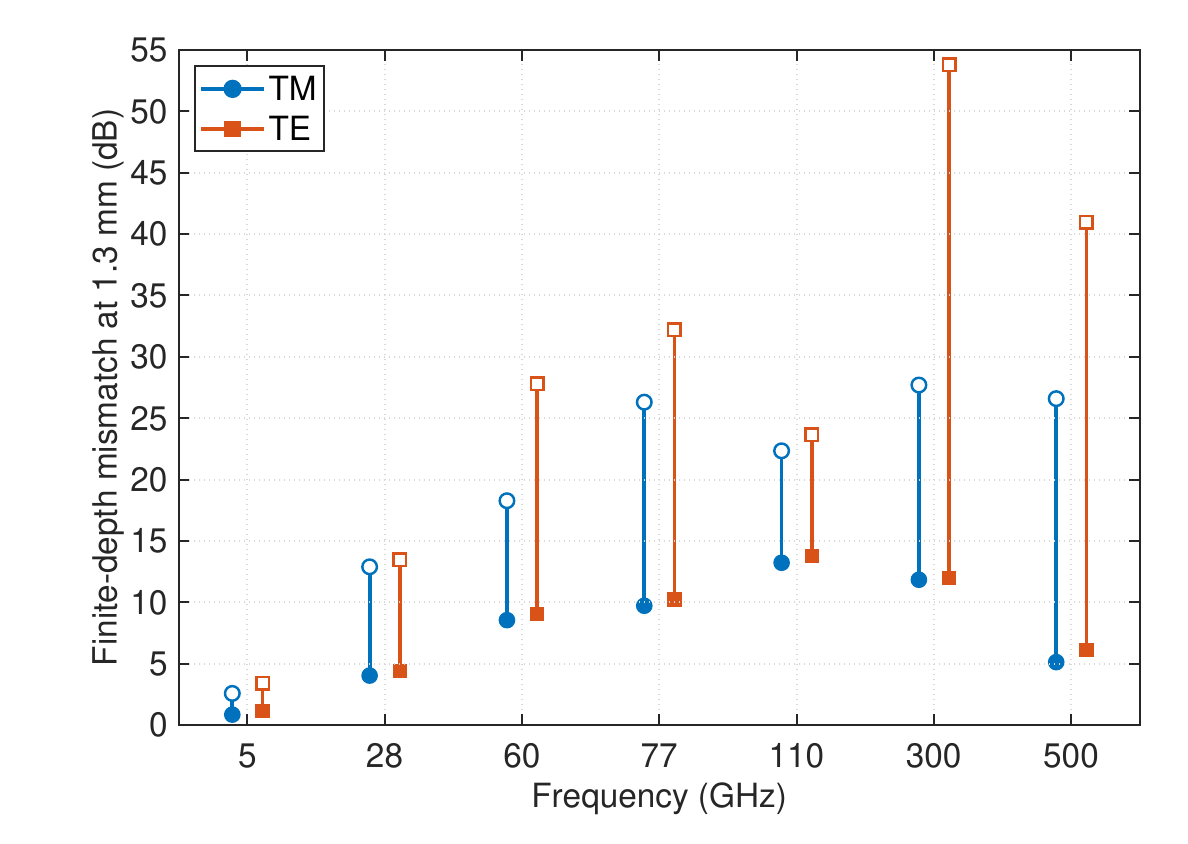}
\caption{Mismatch at 1.3~mm for sampled candidates whose surface mismatch is within $\Delta_{\mathrm{surf}}=0.05~\mathrm{dB}$ of the corresponding polarization-specific optimum. The retained permittivities are held fixed between depths. Filled and open markers show the smallest and largest mismatch at $d=1.3~\mathrm{mm}$ across the sampled permittivity plotting windows.}
\label{fig:surface_nonuniqueness}
\end{figure}

For each strategy, the worst-case mismatches are evaluated separately for TE and TM, with the larger value presented in Fig.~\ref{fig:strategy_comparison}. The centroid provides a common permittivity for both polarizations, while the best single reference and the robust target are selected independently for TE and TM. This comparison complements the joint TE/TM optimization in Table~\ref{tab:headline}. The mathematical formulations are provided in Section~\ref{subsec:benchmark_strategies}. For each polarization, the exact minimax value cannot exceed the mismatch of either admissible benchmark. The resulting margin is particularly relevant at 60~GHz, where optimization reduces the mismatch from 12.23~dB for the best single reference to 9.01~dB. A similar benefit is observed at 77~GHz, with mismatches of 12.04, 16.37 and 10.21~dB for the centroid, the best single reference and robust inverse design, respectively. These differences reflect the nonlinear dependence of interface transmission and attenuation on complex permittivity. Averaging permittivities therefore does not generally minimize the largest deviation in channel loss. The choice of material selection strategy can also determine whether a prescribed tolerance is satisfied. For example, the centroid (4.52~dB) and robust inverse design (3.86~dB) meet the illustrative 6~dB tolerance in this comparison at 500~GHz, while the best single reference yields 6.64~dB. Supplementary Fig.~S4 provides the full range of mismatches obtained from individual skin reference models. A further question concerns the relationship between agreement at the air--skin interface and channel-loss fidelity at finite observation depths. Candidate permittivities are sampled within prescribed plotting windows, retaining those whose polarization-specific surface mismatch lies within $\Delta_{\mathrm{surf}}=0.05~\mathrm{dB}$ of the computed surface optimum. The retained candidates are then evaluated at $d=1.3~\mathrm{mm}$ using the same permittivities (Fig.~\ref{fig:surface_nonuniqueness}). Despite their similar surface mismatch, the sampled TE candidates within the investigated permittivity windows produce mismatches at 1.3~mm of 11.98--53.78~dB at 300~GHz and 6.11--40.95~dB at 500~GHz. This spread arises because similar interface transmission does not uniquely determine attenuation within skin. It demonstrates why phantom acceptance at a given depth must consider both interface transmission and propagation to the specified depth.

\FloatBarrier
\FloatBarrier
\subsection{Sensitivity to reference models and boundary conditions}
\label{subsec:evidence_sensitivity}

\begin{figure}[!htbp]
\centering
\includegraphics[width=0.90\columnwidth]{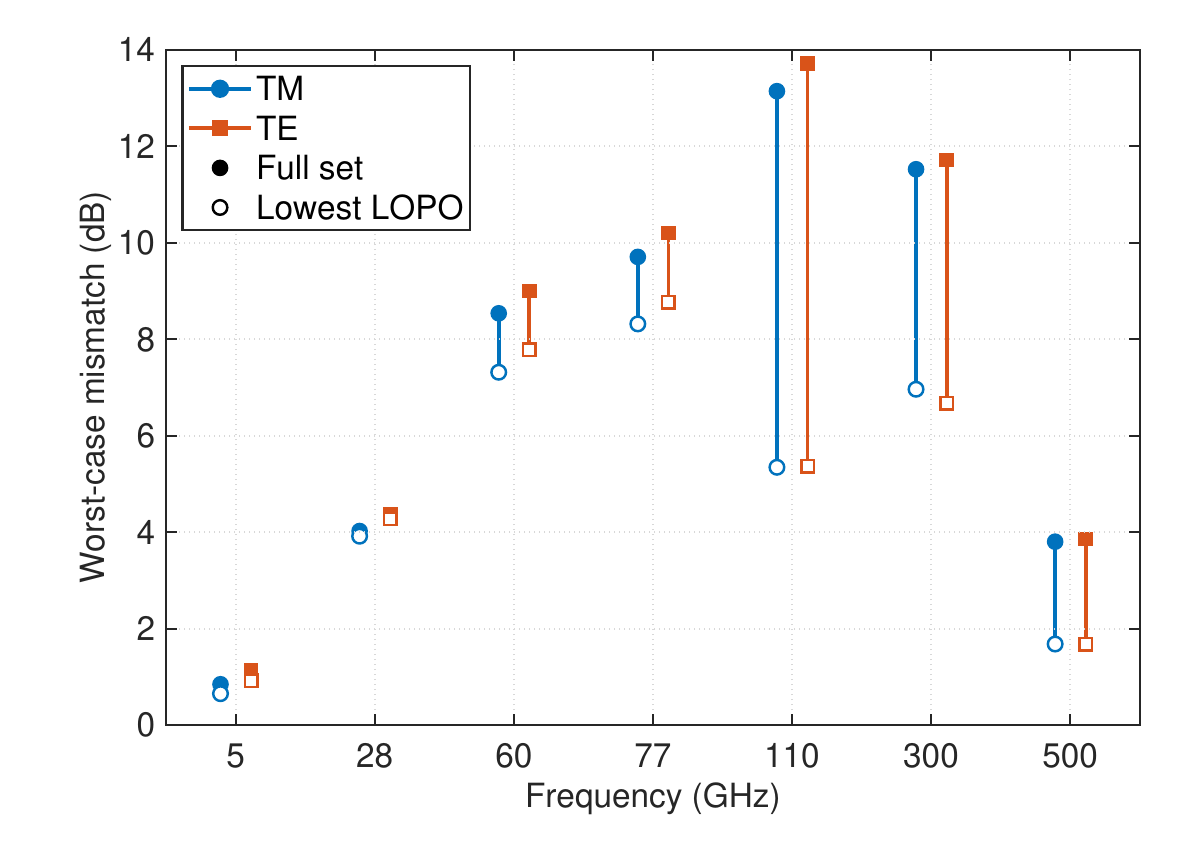}
\caption{LOPO sensitivity at $d=1.3~\mathrm{mm}$. Filled markers represent the separately optimized TE and TM mismatches for the full reference set, while open markers indicate the lowest mismatch after omitting one provenance group. Vertical segments quantify the largest reduction for each frequency and polarization. The corresponding provenance groups are listed in Supplementary Table~S1.}
\label{fig:lopo}
\end{figure}

To examine the influence of the original data on material selection, models sharing measurements or fitted parameters are assigned to provenance groups (cf. Section~\ref{sec:biological_evidence}). Following the principle of leave-one-group-out analysis, a study-specific leave-one-provenance-group-out (LOPO) sensitivity analysis omits each group in turn and repeats the minimax optimization separately for TE and TM at $d=1.3~\mathrm{mm}$. Changes in the resulting minimax mismatch quantify the influence of individual source groups on achievable channel-loss fidelity (cf. Fig.~\ref{fig:lopo}). Since omitting a group reduces the number of reference constraints, the exact minimax optimum can only decrease or remain unchanged.

The influence of individual provenance groups is particularly pronounced at 110 and 300~GHz. Excluding the group associated with the in vivo skin description of Pickwell et al.~\cite{Pickwell2004} reduces the TE minimax mismatch from 13.72 to 5.36~dB at 110~GHz. A similar effect is observed at 300~GHz, where omission of the Chan and Ramer skin reference based on the double-Debye parameters of Pickwell et al.~\cite{Pickwell2005,Chan2018} reduces the corresponding mismatch from 11.72 to 6.67~dB. Across the 14 combinations of frequency and polarization, the largest reductions following removal of one provenance group range from 0.09 to 8.36~dB. Exclusion of the Wallace/Truong group decreases the TE mismatch from 3.86 to 1.68~dB at 500~GHz, while the influence of individual groups is smaller at 28~GHz. These results identify the source groups that most strongly constrain the material specification and illustrate the importance of transparent reference selection. Supplementary Table~S1 summarizes the largest reductions, Table~S16 lists all removals, and Table~S2 examines the admission rules. To examine the influence of reference selection on a common target for both polarizations, the joint TE/TM optimization is repeated at 300~GHz for smaller reference sets at $d=1.3~\mathrm{mm}$. Excluding the Chan model reduces the candidate worst-case mismatch from 11.72 to 6.96~dB, which remains above the illustrative 6~dB budget. For a set containing only the in vivo skin reference from Pickwell et al.~\cite{Pickwell2004} and the healthy-skin values attributed to Zaytsev by Sasaki et al.~\cite{Zaytsev2015APL,Sasaki2022}, the candidate mismatch is 2.05~dB. For each candidate, the gap to the independent analytical lower bound is below 0.001~dB (Supplementary Table~S13). The five-model reference set remains the basis for the primary material-acceptance results.

\begin{figure}[!t]
\centering
\begin{subfigure}{0.93\columnwidth}
\centering
\includegraphics[width=\linewidth]{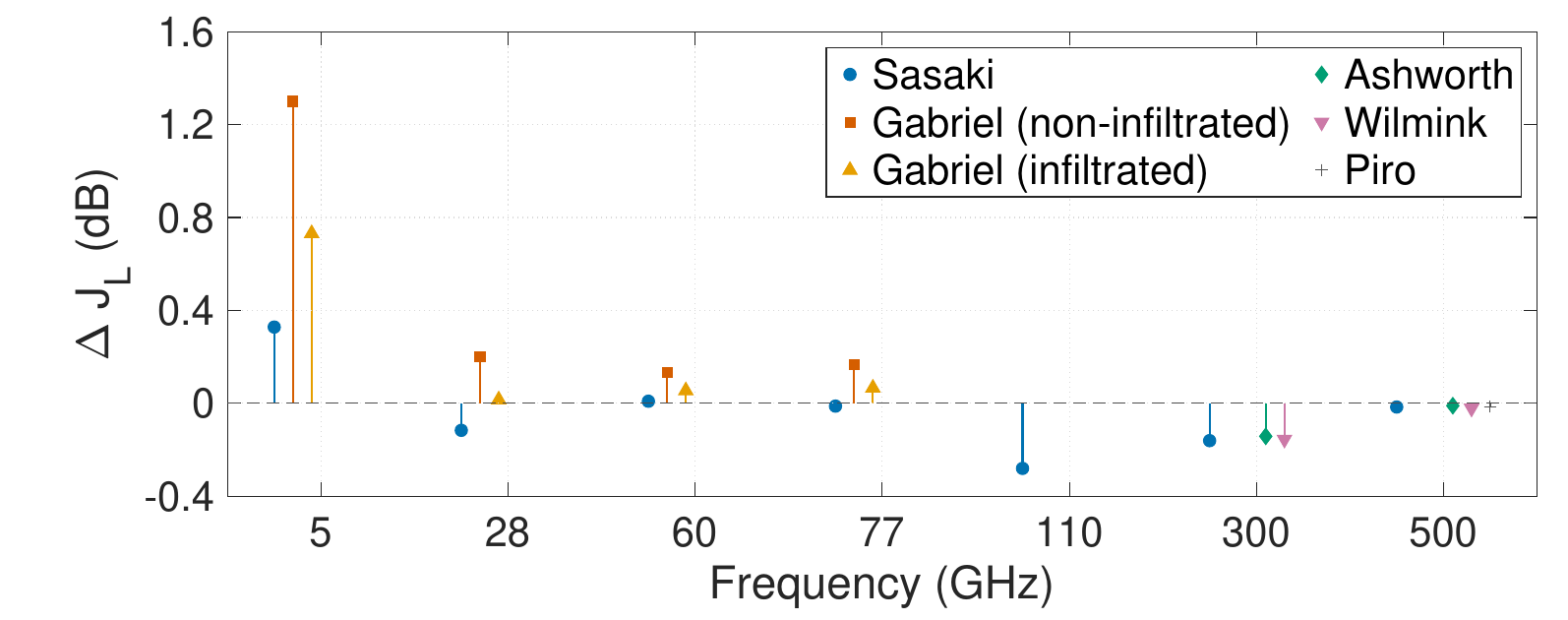}
\caption{TE polarization.}
\label{fig:sat_te}
\end{subfigure}

\vspace{1mm}

\begin{subfigure}{0.93\columnwidth}
\centering
\includegraphics[width=\linewidth]{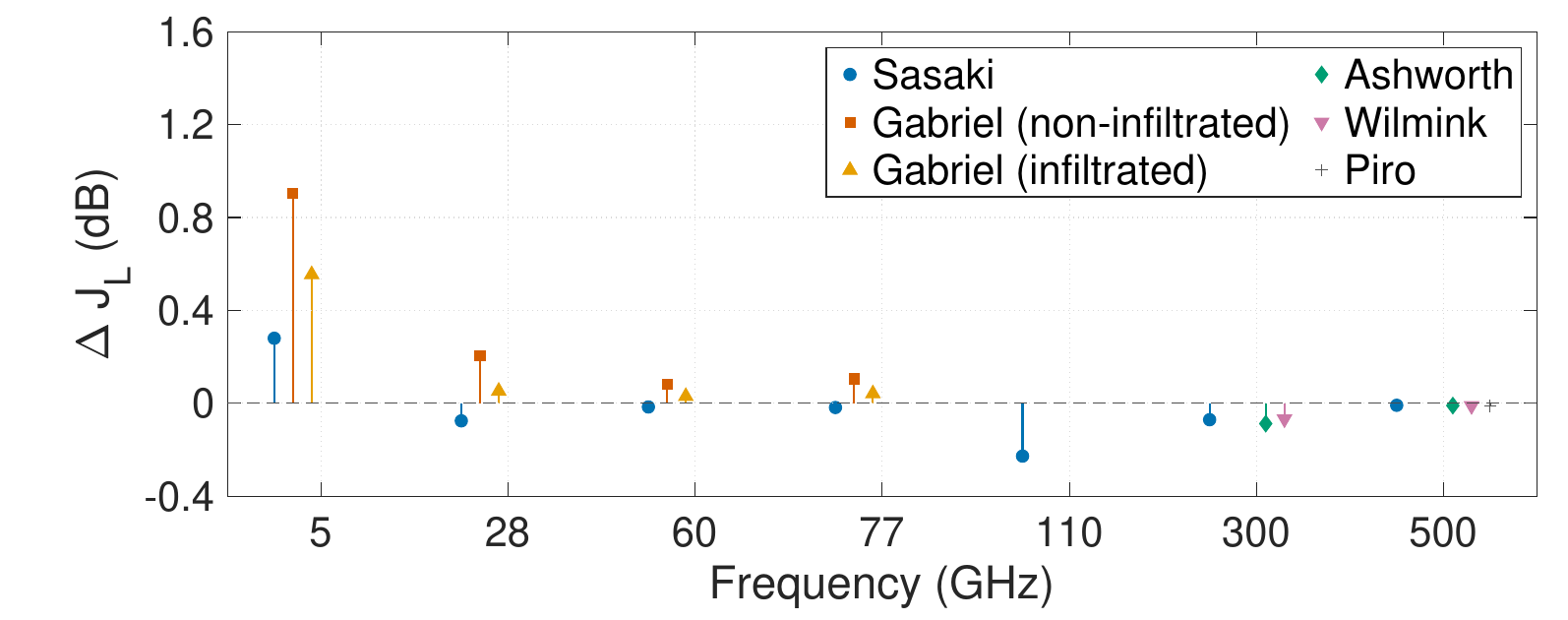}
\caption{TM polarization.}
\label{fig:sat_tm}
\end{subfigure}
\caption{Influence of the skin--SAT boundary on the polarization-specific minimax channel-loss mismatch at $d=1.3~\mathrm{mm}$ for \textbf{(a)} TE and \textbf{(b)} TM polarization. Each marker represents an independent optimization for a specified adipose tissue description. The difference $\Delta J_{\mathrm{L}}=J_{\mathrm{L,skin+SAT}}-J_{\mathrm{L,skin}}$ compares computed minimax candidate values for Case B and Case A, respectively. Positive values indicate a larger mismatch in Case B. Horizontal marker offsets distinguish material descriptions at the same carrier frequency.}
\label{fig:sat}
\end{figure}

The next analysis examines the influence of incidence angles on phantom design. To this end, the maximum incidence angle $\Theta_{\max}$ is successively increased from $0^\circ$ to $30^\circ$, $60^\circ$ and $85^\circ$, using a step size of $1^\circ$. Each range is evaluated separately for TE and TM. At 300~GHz, increasing the maximum incidence angle from $0^\circ$ to $85^\circ$ raises the separately optimized TE mismatch from 10.49 to 11.72~dB and the TM mismatch from 10.49 to 11.53~dB. As each larger angle set includes all previously considered angles, the exact minimax value cannot decrease with increasing angular range. In homogeneous isotropic skin, the attenuation along the surface normal depends on the incidence angle but is identical for TE and TM polarization.
The polarization-dependent differences arise from power transmission at the air--skin interface. Differences between the polarization-specific results arise from polarization-dependent power transmission at the air--skin interface. Agreement at normal incidence alone therefore does not ensure that the same material satisfies the prescribed tolerance over a wider angular range. Supplementary Figs.~S5--S6 trace these changes for all seven carrier frequencies. At an observation depth of 1.3~mm, the largest reference-loss half-range occurs at $85^\circ$ for TE polarization at all seven frequencies (cf. Supplementary Table~S6). The prescribed maximum incidence angle therefore forms part of the material specification. Comparing Case A with Case B separately reveals the influence of the underlying adipose tissue at $d=1.3~\mathrm{mm}$ (cf. Figs.~\ref{fig:framework} and~\ref{fig:sat}). For each of 20 material--frequency combinations of adipose tissue, the same SAT permittivity is assigned to all skin references and phantom candidates, with the skin permittivity optimized separately for TE and TM. The porcine subcutaneous tissue measurements of Sasaki et al.~\cite{Sasaki2017} provide a nominal SAT description across all seven frequencies. Additional models of adipose tissue provide comparisons at selected frequencies, with Sasaki as the sole evaluated model at 110~GHz (cf. Section~\ref{subsec:multilayer_model}).

The influence of the underlying adipose tissue is most pronounced at 5~GHz, where the TE mismatch increases from 1.13~dB in Case A to values between 1.45 and 2.43~dB for the three investigated adipose tissue models. For the remaining six frequencies, the absolute difference between the polarization-specific minimax mismatches of Cases A and B remains below 0.29~dB across all evaluated models. Introducing the skin--SAT interface modifies the input wave admittance seen from air and gives rise to coherent internal reflections within the skin layer. Negative values in Fig.~\ref{fig:sat} indicate closer agreement between candidate and reference losses in Case B, irrespective of whether the absolute channel loss increases or decreases. Supplementary Fig.~S7 and Tables~S7 and~S15 document the dielectric inputs and corresponding optimization results.

\FloatBarrier
\FloatBarrier
\subsection{Numerical verification and certification of feasibility limits}
\label{subsec:verification}

The mismatch of a computed candidate provides an upper bound on the minimax value (cf. Section~\ref{subsec:robust_specification}). A complementary, solver-independent analytical lower bound is derived from half the maximum range of predicted skin losses over the prescribed incidence angles and polarizations (cf. Section~\ref{subsec:feasibility}). Their difference bounds the remaining possible reduction in mismatch. At $d=1.3~\mathrm{mm}$, the largest difference among the joint TE/TM results is $2.12\times10^{-4}$~dB (cf. Supplementary Fig.~S1). Across all seven frequencies and 131 observation depths per frequency, the resulting 917 evaluations satisfy the prescribed optimization tolerance of 0.005~dB.

Determination of the critical depths combines certified feasibility throughout the interval from the skin surface to a lower depth bound with an analytical proof of infeasibility at the upper depth bound. This encloses the first depth at which the tolerance cannot be met, without assuming monotonicity with depth. All finite intervals containing the critical depth are narrower than 0.001~mm (Section~\ref{subsec:feasibility}). Re-evaluating the optimized candidates on a finer $0.01^\circ$ angular grid changes their worst-case mismatch by less than $10^{-4}$~dB. Together, these checks establish the numerical accuracy of the minimax values and critical-depth estimates within the prescribed model and discrete angular constraints. The analytical bounds distinguish limitations imposed by the selected reference losses and design class from limitations of the numerical search, providing a mathematically justified basis for phantom material selection.

\section{Discussion}
\label{sec:discussion}

The principal contribution of the robust inverse design is to translate the allowable error in the skin contribution to link loss into a dielectric material specification with certified feasibility limits. It complements nominal dielectric matching by identifying suitable phantom permittivities and establishing when the required agreement is unattainable within the homogeneous design class. This supports the need for benchmark phantoms and shared validation data~\cite{Skos2026}. Through the acceptance set, the requirements of the experiment directly guide material selection. A nonempty set guides the selection or development of a material with suitable permittivity, subject to the constraints of the fabrication process. An empty set identifies specifications that require a change in the experimental requirements or the material class.

Comparison with the benchmarks clarifies the advantage of optimizing the propagation response itself. While the centroid assigns equal weight to each source group, the minimax criterion limits the largest loss difference without assigning probabilities to the references. Even the best single reference is selected against the full set, so its larger mismatch reflects the restriction to published permittivities. The surface-matching comparison reveals a complementary limitation: similar interface transmission can coexist with markedly different attenuation within skin. Both contributions to loss therefore determine the suitability of a phantom at the intended observation depth. Each resulting target is specific to its frequency and observation depth. Consequently, phantom design for multiple depths or broadband operation requires additional constraints on the common permittivity or its dispersion, respectively. Evaluation of published phantom materials demonstrates how a fixed complex permittivity can be assessed at different depths (cf. Section~\ref{subsec:published_phantom}), with sensitivity to material parameters estimated from published spectra examined in Supplementary Table~S12. Material acceptance based on measured permittivity must account for characterization uncertainty. Hydration-dependent dielectric response and surface texture can also affect phantom measurements~\cite{Mizuno2024,Yamazaki2025,Saxena2024}.

The achievable channel-loss fidelity is closely linked to the selection of skin reference models, as illustrated by the LOPO analysis and the comparison of reference subsets at 300~GHz (cf. Section~\ref{subsec:evidence_sensitivity} and Supplementary Table~S13). Although reducing the reference set can improve attainable agreement, its biological representativeness must be established independently. The reference sets at different frequencies do not provide a population-wide ranking of skin variability. Documenting their origin and frequency coverage is therefore essential for interpreting and reproducing the certified feasibility limits. At the modeled skin thickness, the Case B comparison indicates that the influence of the skin--SAT interface on the optimized mismatch is comparatively small from 28 to 500~GHz for the evaluated adipose tissue descriptions. Attenuation within skin suppresses the contribution of coherent round trips at higher frequencies, although the skin--SAT interface still affects transmitted power. The separate optimizations quantify achievable fidelity for each SAT model. A common phantom for all SAT models would require a joint optimization. Human breast fat and porcine subcutaneous tissue serve as alternative boundary materials for sensitivity analysis. A representative model of human forearm SAT would require corresponding tissue data. Further tissue layers, antenna coupling and propagation within SAT require additional modeling.

Under otherwise identical link conditions, a loss tolerance of 3~dB limits the corresponding deviation in received SNR to 3~dB. For a specific implantable wireless link, the propagation loss investigated in this study can be incorporated into the overall link budget. The angular range and the illustrative loss tolerances of 1, 3 and 6~dB can be adapted to the requirements of the intended experiment. Future studies could extend the framework to broadband and multilayer phantom design, incorporating material dispersion and fabrication constraints into the optimization. Experimental characterization of phantoms designed according to the derived acceptance criteria would further establish their applicability to implantable wireless link evaluation. A practical workflow begins with the frequency, observation depth, angular and polarization coverage, and allowable loss mismatch of the intended experiment. For a feasible requirement, conservative material acceptance requires the entire permittivity uncertainty region to lie within the acceptance set. This criterion guides material selection and characterization, while an infeasible requirement motivates reconsideration of the tolerance, depth or material class before link testing. The method thereby provides a quantitative basis for reproducible material screening before experimental link evaluation.

\section{Methods}
\label{sec:methods}

\subsection{Evaluation conditions}
\label{subsec:evaluation_framework}

For an intended communication experiment, the inverse dielectric specification is defined through

\begin{equation}
\mathcal{S}
=
\left(
f,\,
d,\,
\Theta,\,
\mathcal{P},\,
\mathcal{M}(f),\,
\delta_{\mathrm{L}}
\right),
\label{eq:spec}
\end{equation}

where $f$ denotes the carrier frequency and $d$ the observation depth measured normally from the air--skin interface. The set $\Theta$ specifies the incidence angles, and $\mathcal{P}\subseteq\{\mathrm{TE},\mathrm{TM}\}$ denotes the selected polarizations. At frequency $f$, $\mathcal{M}(f)$ denotes the set of skin reference models, and $\delta_{\mathrm{L}}$ specifies the maximum permissible channel-loss mismatch. Loss tolerances of 1, 3 and 6~dB are considered to investigate increasingly permissive requirements for phantom fidelity. Owing to the logarithmic relationship between channel loss and power transfer, a prescribed loss tolerance $\delta_{\mathrm{L}}$ constrains the ratio of phantom to reference power transfer to the interval $[10^{-\delta_{\mathrm{L}}/10},10^{\delta_{\mathrm{L}}/10}]$ for every selected reference model, incidence angle and polarization. The three tolerances thus correspond to maximum multiplicative factors of approximately 1.26, 2 and 4, respectively. Seven carrier frequencies at 5, 28, 60, 77, 110, 300 and 500~GHz are evaluated independently based on available dielectric reference data, including THz descriptions~\cite{Pickwell2004,Pickwell2005,Truong2013,Truong2015Skin} and with application context from body-centric propagation studies~\cite{Multifaceted}. Case A covers observation depths from 0 to 1.3~mm, while Case B evaluates power transfer into SAT beneath a skin layer of thickness 1.3~mm (cf. Fig.~\ref{fig:framework}). The dielectric acceptance regions are illustrated at observation depths of 0.1 and 1.3~mm. To investigate phantom fidelity over a broad range of incidence conditions, the angular set $\Theta=\{0^\circ,1^\circ,\ldots,85^\circ\}$ is applied to both TE and TM polarizations. The $1^\circ$ grid spans normal to near-grazing incidence, while a refined grid with a resolution of $0.01^\circ$ is used to assess sensitivity to angular discretization. All evaluation conditions are summarized in Supplementary Table~S3.

\subsection{Selection of skin reference models}
\label{sec:biological_evidence}

The skin reference set $\mathcal{M}(f)$ is constructed from published measurements and models of human skin that provide scalar complex permittivities for homogeneous skin or effective values derived in the source. Each complex permittivity must be applicable to the homogeneous forward model without requiring an additional homogenization assumption. Explicitly layer-resolved skin models and fabricated phantom materials are treated separately (cf. Supplementary Tables~S8 and~S10--S12). For each carrier frequency $f$, the reference set $\mathcal{M}(f)$ is constructed from skin models whose applicability is supported by the measurement, modeling or evaluation range reported in the original publication or by a documented subsequent evaluation at that frequency. The corresponding source records document the basis for inclusion, anatomical site, specimen condition, dielectric formulation, underlying experimental data and provenance group. This procedure establishes a traceable reference set of complex permittivities for each investigated frequency.

At 5, 28, 60, and 77~GHz, published skin dispersion models are complemented by in vivo measurements and fitted descriptions from different anatomical regions~\cite{Gabriel1996,Alekseev2007,Chahat2011,Zhekov2019}. Additional human-skin evidence includes broadband in vivo measurements and Cole--Cole representations of excised skin~\cite{Ghodgaonkar2000,Mirbeik2018,Alabaster2005}. The in vivo response of skin at the palm reported by Ghodgaonkar is evaluated at 28~GHz using the original single-Debye parameters, yielding $\varepsilon_{\mathrm{r}}^\ast=8.01-j\,11.55$~\cite{Ghodgaonkar2000}. Alabaster's fitted dielectric function represents an estimate of the response of fresh skin at $37^\circ\mathrm{C}$ obtained by correcting measurements performed on a single formaldehyde-fixed and dried skin specimen~\cite{Alabaster2005}. Homogeneous forearm descriptions are distinguished from layer-resolved palmar descriptions~\cite{Alekseev2007}. Documented subsequent model evaluations support selected descriptions at additional carrier frequencies~\cite{Wu2015}. Broadband coaxial-probe measurements of wrist skin over 0.5--110~GHz~\cite{Hwang2003} provide the basis for a Cole--Cole description evaluated at 5, 28, 60, 77 and 110~GHz in the present study. The fitted complex permittivity represents the combined dielectric response within the sensing region of the probe. For the higher carrier frequencies, admitted THz descriptions draw on published skin spectroscopy and associated dielectric parameterizations~\cite{Pickwell2004,Pickwell2005,Chan2018,Truong2013}. Their inclusion follows documented experimental or model-evaluation support, with relevant double-Debye fits contributing at 300 and 500~GHz. The normal-skin parameters used for the Truong description follow Table~4.1 of the doctoral thesis and the corresponding peer-reviewed double-Debye study~\cite{TruongThesis2015,Truong2015Skin}. Truong's fitted normal-skin response is linked to the ex-vivo spectroscopy of Wallace et al.~\cite{Wallace2006}, while Chan and Ramer use the parameters for simulated in vivo normal skin from Table~2 of Pickwell et al.~\cite{Pickwell2005,Chan2018}. Both dependencies are represented in the provenance-group assignments.

An additional healthy-skin reference at 300~GHz is obtained from Table~C1 of Sasaki et al., which attributes the real relative permittivity $\varepsilon_{\mathrm{r}}'=3.3$ and electrical conductivity $\sigma=55~\mathrm{S/m}$ to Zaytsev et al.~\cite{Zaytsev2015APL,Sasaki2022}. The corresponding dielectric-loss component is calculated as $\varepsilon_{\mathrm{r}}''=\sigma/(2\pi f\varepsilon_0)$, where $\varepsilon_0$ denotes the vacuum permittivity. This yields $\varepsilon_{\mathrm{r}}^\ast\approx3.30-j\,3.30$ (cf. Supplementary Table~S5). Other anatomical measurements and alternative dispersive or multilayer representations were considered according to the selection criteria for $\mathcal{M}(f)$~\cite{Echchgadda2013,Zaytsev2015,Mattana2024,Christ2025}. Supplementary Table~S8 documents the treatment of additional published descriptions under the present numerical-admission rules. Altogether, the reference data comprise 19 individual dielectric models and 51 entries across the seven carrier frequencies. Models derived from shared measurements or fitted dielectric parameters are assigned to common provenance groups, providing the basis for the LOPO sensitivity analysis and the centroid benchmark. Information on anatomical site, specimen preparation and reported temperature is retained in the corresponding source records. Figure~\ref{fig:evidence} summarizes the source families, and Supplementary Table~S4 lists the individual descriptions, frequencies and provenance groups. All individual descriptions enter the primary maximum. For the 300~GHz subset comparison at 1.3~mm, the complete five-model set is compared with a four-model set excluding Chan and a two-model set retaining the in vivo skin reference from Pickwell et al.~\cite{Pickwell2004} and the Zaytsev healthy-skin reference~\cite{Zaytsev2015APL,Sasaki2022}. Each set is optimized jointly and separately for TE and TM using the original angular grid (cf. Supplementary Table~S13).

\subsection{Air--skin interface and propagation within skin}
\label{subsec:skin_forward}

The channel loss is determined from the normal component of the time-averaged plane-wave power flux at depth $d$, normalized to the incident normal power flux at the air--skin interface. For Case A, the propagation model combines interface transmission and attenuation during propagation~\cite{BalanisAEM2012}. Consider the interface at $z=0$, with lossless air occupying $z<0$ and homogeneous skin extending into $z>0$. The $z$-axis extends into the tissue along the interface normal, while the $x$-axis lies within the plane of incidence parallel to the interface. The incidence angle $\Theta_{\mathrm{i}}$ is measured from the interface normal, with the observation plane located at $z=d$. Using the $e^{j2\pi ft}$ phasor convention, where $t$ denotes time and $j^2=-1$, the complex permittivity of skin reference $m\in\mathcal{M}(f)$ is expressed as

\begin{equation}
\varepsilon_m^{\ast}
=
\varepsilon_m'
-
j\varepsilon_m'',
\qquad
\varepsilon_m''>0,
\label{eq:complex_permittivity}
\end{equation}

where $\varepsilon_m'=\operatorname{Re}\{\varepsilon_m^\ast\}$ denotes the real part and $\varepsilon_m''=-\operatorname{Im}\{\varepsilon_m^\ast\}$ denotes the positive dielectric-loss component. The frequency dependence of $\varepsilon_m^\ast$ is implicit in the following expressions. Continuity of the tangential electric and magnetic fields at the air--skin interface gives the Fresnel reflection coefficients~\cite{Wu2015,BalanisAEM2012}

\begin{align}
\Gamma_{\mathrm{TE},m}(\Theta_{\mathrm{i}})
&=
\frac{
\cos\Theta_{\mathrm{i}}
-
\sqrt{\varepsilon_m^\ast-\sin^2\Theta_{\mathrm{i}}}
}{
\cos\Theta_{\mathrm{i}}
+
\sqrt{\varepsilon_m^\ast-\sin^2\Theta_{\mathrm{i}}}
},
\label{eq:fresnel_te}
\\
\Gamma_{\mathrm{TM},m}(\Theta_{\mathrm{i}})
&=
\frac{
\varepsilon_m^\ast\cos\Theta_{\mathrm{i}}
-
\sqrt{\varepsilon_m^\ast-\sin^2\Theta_{\mathrm{i}}}
}{
\varepsilon_m^\ast\cos\Theta_{\mathrm{i}}
+
\sqrt{\varepsilon_m^\ast-\sin^2\Theta_{\mathrm{i}}}
}.
\label{eq:fresnel_tm}
\end{align}

Because the incident medium is lossless, conservation of the normal time-averaged power flux at the interface gives the fraction entering the lossy skin half-space as

\begin{equation}
R_{p,m}
=
\left|\Gamma_{p,m}\right|^2,
\qquad
T_{p,m}
=
1-R_{p,m},
\qquad
p\in\{\mathrm{TE},\mathrm{TM}\}.
\label{eq:RT}
\end{equation}

The quantity $R_{p,m}$ denotes the fraction of incident power reflected at the air--skin interface, while $T_{p,m}$ denotes the fraction transmitted into the skin, for skin reference model $m$ and polarization $p$. Beyond the interface, attenuation is governed by the complex normal wavenumber. Conservation of the tangential wavenumber gives

\begin{equation}
k_x
=
k_0\sin\Theta_\mathrm{i},
\end{equation}

where $k_0=2\pi f/c_0$ and $c_0$ is the speed of light in vacuum. For the nonmagnetic lossy media considered in this study, the corresponding normal wavenumber is~\cite{BalanisAEM2012}

\begin{equation}
k_{z,m}
=
k_0
\sqrt{
\varepsilon_m^\ast-\sin^2\Theta_{\mathrm{i}}
},
\label{eq:kz}
\end{equation}

where the square-root branch is selected with $\operatorname{Re}\{k_{z,m}\}\geq0$ and $\operatorname{Im}\{k_{z,m}\}\leq0$ to ensure propagation into the tissue with non-increasing field amplitude. Under the spatial dependence $e^{-j(k_xx+k_{z,m}z)}$, the normal field-attenuation coefficient is

\begin{equation}
\alpha_{z,m}
=
-\operatorname{Im}\{k_{z,m}\}
\geq 0.
\label{eq:alpha_z}
\end{equation}

Because power is quadratic in field amplitude, the normal power flux decreases with twice the field-attenuation exponent. The propagation factor at depth $d$ is

\begin{equation}
L_{\mathrm{prop},m}
(f,\Theta_{\mathrm{i}},d)
=
\exp\left(-2\alpha_{z,m}d\right).
\label{eq:propagation_factor}
\end{equation}

Combining this factor with the transmitted power fraction at the air--skin interface gives

\begin{equation}
\mathcal{T}_{p,m}(f,d,\Theta_\mathrm{i})
=
T_{p,m}
L_{\mathrm{prop},m}(f,\Theta_\mathrm{i},d).
\label{eq:power}
\end{equation}

This dimensionless quantity is the normal power flux at depth $d$, normalized to the incident normal flux at the interface. It assumes semi-infinite skin and excludes reflections from a deeper tissue boundary, which are treated separately. Its logarithmic form defines the channel-loss metric used throughout the inverse design

\begin{equation}
L_{\mathrm{ch},p}^{(m)}
(f,d,\Theta_{\mathrm{i}})
=
-10\log_{10}
\mathcal{T}_{p,m}
(f,d,\Theta_{\mathrm{i}}).
\label{eq:loss}
\end{equation}

The channel-loss metric separates into an interface contribution, $-10\log_{10}(1-R_{p,m})$, and a depth-dependent contribution, $(20/\ln 10)\alpha_{z,m}d$. For a fixed complex permittivity of a homogeneous isotropic medium and incidence angle, the normal attenuation coefficient $\alpha_{z,m}$ is identical for TE and TM polarization. The polarization dependence of the channel loss therefore originates from the Fresnel power-transmission factor. At the skin surface ($d=0$), only the interface loss contributes, while propagation into the tissue introduces additional attenuation governed by the imaginary part of the normal wavenumber. Applying the plane-wave propagation model derived above to a phantom candidate with complex permittivity $\varepsilon_{\mathrm{c}}^\ast$ yields the corresponding channel loss $L_{\mathrm{ch},p}^{(\mathrm{c})}$. The difference between the candidate and reference skin losses can be expressed as

\begin{equation}
\begin{aligned}
L_{\mathrm{ch},p}^{(\mathrm{c})}
-L_{\mathrm{ch},p}^{(m)}
={}&-10\log_{10}
\left(\frac{T_{p,\mathrm{c}}}{T_{p,m}}\right)+\frac{20d}{\ln 10}
\left(\alpha_{z,\mathrm{c}}-\alpha_{z,m}\right).
\end{aligned}
\label{eq:loss_decomposition}
\end{equation}

In this expression, $T_{p,\mathrm{c}}$ and $\alpha_{z,\mathrm{c}}$ denote the interface power-transmission coefficient and the normal field-attenuation coefficient of the phantom candidate, respectively. The first term quantifies the interface-transmission difference, while the second describes the depth-dependent attenuation difference. This decomposition explains why a channel-loss mismatch can persist at the surface and why nearly equivalent surface responses can produce substantially different losses at finite depths. Applying an identical propagation model to all skin reference models and candidate permittivities provides the common basis for the inverse formulation of Section~\ref{sec:inverse_design}. The propagation model is subsequently extended to a finite skin layer above SAT in Case B.

\subsection{Finite skin layer with subcutaneous adipose tissue}
\label{subsec:multilayer_model}

Case B replaces the skin half-space (Case A, Section~\ref{subsec:skin_forward}) with a skin layer of thickness $d_{\mathrm{skin}}=1.3~\mathrm{mm}$ backed by semi-infinite SAT (Fig.~\ref{fig:framework}). The nominal SAT model is based on complex-permittivity spectra of porcine subcutaneous tissue measured in vitro by Sasaki et al.~\cite{Sasaki2017} at approximately $35~^\circ\mathrm{C}$ over a frequency range from 1 to 1000~GHz, covering all seven investigated frequencies. To assess sensitivity to alternative adipose tissue descriptions, the original models for blood-infiltrated and non-blood-infiltrated fat are evaluated at 5, 28, 60 and 77~GHz~\cite{Gabriel1996}. The investigation at 300 and 500~GHz includes published spectra of human breast adipose tissue and porcine adipose tissue~\cite{Ashworth2009,Wilmink2011}. An additional Havriliak--Negami model is considered at 500~GHz, the only investigated frequency within its reported fitting range of 0.5 to 1.5~THz~\cite{Piro2016}. Every input is evaluated within the reported measurement or model range. Spectral values reconstructed from published figures are approximate material inputs, as documented in Supplementary Table~S7. For each investigated SAT model, an identical complex permittivity is assigned to the underlying adipose tissue in all skin reference and phantom candidate configurations. Independent polarization-specific optimizations then isolate the sensitivity of achievable channel-loss fidelity to the prescribed lower boundary. Both cases include reflection at the air--skin interface. Case B additionally includes reflection at the skin--SAT interface and repeated reflection within the finite skin layer. We account for these coherent multiple reflections with the characteristic-matrix method for stratified media~\cite{YehLayeredMedia}. For each nonmagnetic medium $\ell\in\{\mathrm{air},\mathrm{skin},\mathrm{SAT}\}$, the polarization-dependent wave admittances are

\begin{equation}
Y_{\ell,\mathrm{TE}}=\frac{k_{z,\ell}}{2\pi f\mu_0},
\qquad
Y_{\ell,\mathrm{TM}}=
\frac{2\pi f\varepsilon_0\varepsilon_\ell^\ast}{k_{z,\ell}},
\label{eq:wave_admittance}
\end{equation}

where $\varepsilon_\ell^\ast$ denotes the complex relative permittivity of medium $\ell$ (with $\varepsilon_{\mathrm{air}}^\ast=1$), and $k_{z,\ell}$ follows from the conserved tangential wavenumber using the square-root branch defined in Eq.~\eqref{eq:kz}. With $\varepsilon_0$ and $\mu_0$ denoting the vacuum
permittivity and permeability, respectively, the normalized admittances are defined as $\widetilde{Y}_{\ell,p}=Y_{\ell,p}/Y_{\mathrm{vac}}$,
where $Y_{\mathrm{vac}}=\sqrt{\varepsilon_0/\mu_0}$ is the vacuum admittance. Using a common normalization leaves the reflection coefficient and normalized power transfer unchanged. For a wave propagating towards increasing $z$, the signed tangential fields satisfy $H_{\mathrm{t}}=Y_{\ell,p}E_{\mathrm{t}}$. We use $(E_{\mathrm{t}},H_{\mathrm{t}})=(E_y,-H_x)$ for TE and $(E_x,H_y)$ for TM polarization. With these conventions, the characteristic matrix relates the total tangential fields, including forward and backward waves, at the two boundaries of the skin layer:

\begin{equation}
\begin{bmatrix}
E_{\mathrm{t}}(0)\\
H_{\mathrm{t}}(0)
\end{bmatrix}
=
\mathbf{M}_{\mathrm{skin},p}
\begin{bmatrix}
E_{\mathrm{t}}(d_{\mathrm{skin}})\\
H_{\mathrm{t}}(d_{\mathrm{skin}})
\end{bmatrix}.
\label{eq:matrix_definition}
\end{equation}

For polarization $p$, the transfer matrix $\mathbf{M}_{\mathrm{skin},p}$ relates the tangential field components at the skin--SAT interface to those at the air--skin interface. Using the $e^{j2\pi ft}$ time convention and the spatial propagation factor $e^{-jk_{z,\mathrm{skin}}z}$, this matrix is expressed as

\begin{equation}
\mathbf{M}_{\mathrm{skin},p}
=
\begin{bmatrix}
\cos\delta_{\mathrm{skin}}
&
j\sin\delta_{\mathrm{skin}}/Y_{\mathrm{skin},p}
\\
jY_{\mathrm{skin},p}\sin\delta_{\mathrm{skin}}
&
\cos\delta_{\mathrm{skin}}
\end{bmatrix},
\label{eq:skin_matrix}
\end{equation}

with

\begin{equation}
\delta_{\mathrm{skin}}
=
k_{z,\mathrm{skin}}d_{\mathrm{skin}}.
\label{eq:skin_phase}
\end{equation}

The product $\delta_{\mathrm{skin}}=k_{z,\mathrm{skin}}d_{\mathrm{skin}}$ describes propagation across the skin layer. Its real part represents the electrical length along the interface normal, while its negative imaginary part determines the attenuation exponent of the field amplitude. The dependence on the incidence angle is incorporated through $k_{z,\mathrm{skin}}$. Within semi-infinite SAT, only the wave propagating away from the skin--SAT interface is retained. The radiation condition therefore gives $H_{\mathrm{t}}(d_{\mathrm{skin}}) =Y_{\mathrm{SAT},p}E_{\mathrm{t}}(d_{\mathrm{skin}})$. Substitution into~\eqref{eq:matrix_definition} gives the input admittance seen from air:

\begin{equation}
Y_{\mathrm{in},p}
=
\frac{
M_{21}+M_{22}Y_{\mathrm{SAT},p}
}{
M_{11}+M_{12}Y_{\mathrm{SAT},p}
},
\label{eq:input_admittance}
\end{equation}

where $M_{ij}$ abbreviates the $(i,j)$ element of $\mathbf{M}_{\mathrm{skin},p}$ for the polarization $p$ under evaluation. The reflection coefficient at the air--skin boundary of Case B then follows as

\begin{equation}
\Gamma_{\mathrm{multi},p}=
\frac{Y_{\mathrm{air},p}-Y_{\mathrm{in},p}}
{Y_{\mathrm{air},p}+Y_{\mathrm{in},p}}.
\label{eq:multilayer_reflection}
\end{equation}

This reflection coefficient refers to the signed tangential electric-field amplitude. For TM polarization its sign differs from the convention used in~\eqref{eq:fresnel_tm}, but its squared magnitude is unchanged. The field-transmission coefficient from the incident wave in air to the outgoing wave at the skin--SAT interface is

\begin{equation}
t_{\mathrm{multi},p}=
\frac{2Y_{\mathrm{air},p}}
{Y_{\mathrm{air},p}(M_{11}+M_{12}Y_{\mathrm{SAT},p})
+M_{21}+M_{22}Y_{\mathrm{SAT},p}}.
\label{eq:multilayer_transmission_field}
\end{equation}

For peak-amplitude phasors, the normal time-averaged power flux of a forward wave is $\tfrac{1}{2}\operatorname{Re}\{Y_{\ell,p}\}|E_{\mathrm{t}}|^2$. The factor $1/2$ cancels when normalizing transmitted to incident power flux. Consequently, the transmitted field amplitude must be weighted by the real wave admittances to obtain the power entering SAT, normalized to incident power in air:

\begin{equation}
\mathcal{T}_{\mathrm{multi},p}
=
\frac{
\operatorname{Re}\{Y_{\mathrm{SAT},p}\}
}{
\operatorname{Re}\{Y_{\mathrm{air},p}\}
}
\left|
t_{\mathrm{multi},p}
\right|^2.
\label{eq:multilayer_power}
\end{equation}

As a consistency limit, setting the SAT permittivity equal to that of the skin removes the dielectric contrast at the lower interface. The layered power-transfer expression then reduces to $\mathcal{T}_{p,m}(f,d_{\mathrm{skin}},\Theta_{\mathrm{i}})=(1-R_{p,m})\exp(-2\alpha_{z,m}d_{\mathrm{skin}})$, recovering the homogeneous forward response of Eq.~\eqref{eq:power}. Expressing the normalized power-transfer factor in decibels yields the layered channel-loss metric

\begin{equation}
L_{\mathrm{ch},p}^{\mathrm{multi}}
=
-10\log_{10}\mathcal{T}_{\mathrm{multi},p}.
\label{eq:multilayer_loss}
\end{equation}

Case B includes reflection at both interfaces, absorption in skin and coherent multiple reflections. At $d=d_{\mathrm{skin}}$, Case A predicts forward power within skin (Eq.~\eqref{eq:loss}), while Case B predicts power entering SAT (Eq.~\eqref{eq:multilayer_loss}). The influence of the skin--SAT interface is assessed
by comparing the optimized mismatches for Cases A and B.

\subsection{Minimax formulation and acceptance criteria}
\label{sec:inverse_design}
\label{subsec:robust_specification}

Under the propagation model of Section~\ref{subsec:skin_forward}, each selected skin reference model and candidate permittivity is mapped to the same power-based channel-loss metric. The reference and candidate are evaluated at identical carrier frequency, depth, incidence angle, and polarization. For candidate complex permittivity $\varepsilon_{\mathrm{c}}^\ast$, the signed loss difference relative to reference $m$ is

\begin{equation}
\begin{split}
\Delta L_{\mathrm{ch},m,p}(\varepsilon_{\mathrm{c}}^\ast;f,d,\Theta_\mathrm{i})
&=L_{\mathrm{ch},p}^{(\mathrm{c})}(f,d,\Theta_\mathrm{i})-L_{\mathrm{ch},p}^{(m)}(f,d,\Theta_\mathrm{i}),
\end{split}
\label{eq:loss_mismatch}
\end{equation}

where $p\in\mathcal{P}$ denotes the polarization state. Its absolute value gives the loss mismatch in dB. A single candidate is assessed jointly against all selected references, sampled incidence angles, and prescribed polarizations through

\begin{equation}
J_{\mathrm{L}}(\varepsilon_{\mathrm{c}}^\ast;f,d)=
\max_{\substack{m\in\mathcal{M}(f)\\
\Theta_\mathrm{i}\in\Theta,\;p\in\mathcal{P}}}
\left|\Delta L_{\mathrm{ch},m,p}(\varepsilon_{\mathrm{c}}^\ast;f,d,\Theta_\mathrm{i})\right|.
\label{eq:JL}
\end{equation}

Thus, $J_{\mathrm{L}}$ quantifies the largest modeled channel-loss mismatch across the selected skin reference models, incidence angles, and prescribed polarizations. At each frequency and observation depth, a single phantom candidate must satisfy all these evaluation conditions simultaneously. This criterion specifies the accuracy of the modeled skin contribution to link loss in Fig.~\ref{fig:framework}. Angular maximization uses the prescribed discrete set $\Theta$. For a consistent SNR comparison at fixed frequency, depth, reference, angle and polarization, abbreviate the signed loss difference as $\Delta L_{\mathrm{ch}}=\Delta L_{\mathrm{ch},m,p}$ and $\Delta\mathrm{SNR}_{\mathrm{dB}}=\mathrm{SNR}_{\mathrm{dB}}^{(\mathrm{c})}-\mathrm{SNR}_{\mathrm{dB}}^{(m)}$. If transmit power, receiver-noise power, and all other link terms are held fixed, then

\begin{equation}
\begin{aligned}
\Delta\mathrm{SNR}_{\mathrm{dB}}
&=-\Delta L_{\mathrm{ch}},\\
\left|\Delta\mathrm{SNR}_{\mathrm{dB}}\right|
&=\left|\Delta L_{\mathrm{ch}}\right|.
\end{aligned}
\label{eq:snr}
\end{equation}

At each frequency $f$, the admissible design class is defined as $\mathcal{D}(f)=\{\varepsilon'-j\varepsilon'':\varepsilon'>0,\,\varepsilon''\geq0\}$. The design class comprises complex relative permittivities of homogeneous dielectric media with positive real and nonpositive imaginary parts. Under the stated phasor convention, the nonnegative loss component enforces passivity at the evaluated frequency. Each frequency is optimized independently without imposing a common broadband dispersion law. The best achievable worst-case mismatch is defined by the infimum

\begin{equation}
J_{\mathrm{L}}^\star(f,d)=
\inf_{\varepsilon_{\mathrm{c}}^\ast\in\mathcal{D}(f)}
J_{\mathrm{L}}(\varepsilon_{\mathrm{c}}^\ast;f,d).
\label{eq:minimax}
\end{equation}

Each inverse problem is defined at one frequency and one prescribed depth, with robustness imposed jointly over $\mathcal{M}(f)$, $\Theta$, and $\mathcal{P}$. For an application-defined channel-loss budget $\delta_{\mathrm{L}}$, the corresponding admissible dielectric region is

\begin{equation}
\Omega_{\delta_{\mathrm{L}}}(f,d)=
\left\{\varepsilon_{\mathrm{c}}^\ast\in\mathcal{D}(f):
J_{\mathrm{L}}(\varepsilon_{\mathrm{c}}^\ast;f,d)\leq\delta_{\mathrm{L}}\right\}.
\label{eq:Omega}
\end{equation}

The set $\Omega_{\delta_{\mathrm{L}}}(f,d)$ contains all candidate permittivities satisfying the prescribed tolerance, allowing published phantom materials
to be evaluated against the same criteria. An empty set indicates infeasibility in the defined material class. Computed objectives and acceptance decisions use the stored numerical material inputs at full precision, independently of rounding for display. Minimization of $J_{\mathrm{L}}$ uses a deterministic grid search with successive local refinement of the real and loss components of the complex permittivity. The initial search domain in the complex-permittivity plane is defined by the minimum and maximum values of both components across the selected skin reference models:

\begin{align}
\min_{m\in\mathcal{M}(f)}\varepsilon_m'
&\leq\varepsilon_{\mathrm{c}}'
\leq\max_{m\in\mathcal{M}(f)}\varepsilon_m',\\
\min_{m\in\mathcal{M}(f)}\varepsilon_m''
&\leq\varepsilon_{\mathrm{c}}''
\leq\max_{m\in\mathcal{M}(f)}\varepsilon_m''.
\label{eq:search_domain}
\end{align}

These bounds define the initial search domain, which can be expanded within the admissible design class $\mathcal{D}(f)$. For joint TE/TM optimization, the initial grid is sampled with a spacing of $0.20$ along both the $\varepsilon_{\mathrm{c}}'$ and $\varepsilon_{\mathrm{c}}''$ axes. Up to ten candidates with the smallest values of $J_{\mathrm{L}}$ are retained for local refinement. Around each retained candidate, a finer grid with a spacing of $0.025$ covers a neighborhood extending two coarse grid steps in each coordinate, clipped to the current search bounds. Two subsequent refinements around the best candidate use spacings of $0.005$ and $0.001$, respectively, with neighborhoods extending two preceding grid steps. Each candidate is evaluated on the $1^\circ$ incidence angle grid prescribed in Section~\ref{subsec:evaluation_framework} and compared with the best value found. Numerical accuracy is assessed by comparing the computed candidate objective, which provides an upper bound on the minimax value, with the analytical lower bound $J_{\mathrm{LB}}$ derived in Section~\ref{subsec:feasibility}. If the difference exceeds 0.005~dB, both search intervals are expanded by their current widths on either side within the admissible design class. Search and refinement continue until the prescribed accuracy is achieved or six expansions have been attempted. If this accuracy is not achieved, the minimax value remains unresolved at that accuracy. Section~\ref{subsec:verification} summarizes the verification results. For visualization of the acceptance sets, each reference permittivity interval is extended by its width on either side within the admissible design class. The resulting fixed plotting domains define the displayed portion of the acceptance sets independently of any expansion of the numerical search domain. Once a dielectric target has been established, the complex permittivity $\varepsilon_{\mathrm{mat}}^\ast\in\mathcal{D}(f)$ of a realized phantom material can be evaluated by its excess mismatch relative to the minimax value:

\begin{equation}
\begin{split}
\Delta J_{\mathrm{design}}(\varepsilon_{\mathrm{mat}}^\ast;f,d)
&=J_{\mathrm{L}}(\varepsilon_{\mathrm{mat}}^\ast;f,d)-J_{\mathrm{L}}^\star(f,d)\geq0.
\end{split}
\label{eq:dJ}
\end{equation}

The design penalty $\Delta J_{\mathrm{design}}$ quantifies the additional worst-case mismatch of a material relative to the exact minimax value. Using the
computed candidate objective instead changes this penalty by no more than the verified optimality gap. A material satisfies the prescribed channel-loss
tolerance if $J_{\mathrm{L}}(\varepsilon_{\mathrm{mat}}^\ast;f,d) \leq\delta_{\mathrm{L}}$.

\subsection{Feasibility and critical depth}
\label{subsec:feasibility}

For a given frequency, depth, incidence angle, and polarization,
the selected skin reference models span an interval of
channel-loss responses. Its endpoints are

\begin{align}
L_{\max,p}(f,d,\Theta_\mathrm{i})
&=\max_{m\in\mathcal{M}(f)}
L_{\mathrm{ch},p}^{(m)}(f,d,\Theta_\mathrm{i}),\\
L_{\min,p}(f,d,\Theta_\mathrm{i})
&=\min_{m\in\mathcal{M}(f)}
L_{\mathrm{ch},p}^{(m)}(f,d,\Theta_\mathrm{i}).
\end{align}

Let $\ell_{\mathrm{c}}$ denote the channel loss predicted for an admissible phantom candidate under the same
conditions. The maximum deviation from the smallest and largest reference losses satisfies

\begin{equation*}
\max\left\{
|\ell_{\mathrm{c}}-L_{\min,p}|,
|\ell_{\mathrm{c}}-L_{\max,p}|
\right\}
\geq
\frac{L_{\max,p}-L_{\min,p}}{2},
\end{equation*}

where the arguments of $L_{\min,p}$ and $L_{\max,p}$ are held fixed. By the triangle inequality, this relation holds for every candidate within the admissible design class $\mathcal{D}(f)$, independently of the optimization algorithm. Taking the largest half-range of the reference losses across all prescribed incidence angles and polarizations yields the analytical lower bound

\begin{equation}
\begin{split}
J_{\mathrm{LB}}(f,d)
&=\frac{1}{2}
\max_{\substack{\Theta_\mathrm{i}\in\Theta\\p\in\mathcal{P}}}
\bigl[L_{\max,p}(f,d,\Theta_\mathrm{i})-L_{\min,p}(f,d,\Theta_\mathrm{i})\bigr].
\end{split}
\label{eq:LB}
\end{equation}

The analytical lower bound need not be attained by a single complex permittivity because the same candidate must satisfy all prescribed angular and polarization
conditions simultaneously. Evaluating any candidate $\widehat{\varepsilon}_{\mathrm{c}}^\ast\in\mathcal{D}(f)$ provides an upper bound, where the hat denotes a
computed phantom candidate. Together, these bounds enclose the exact minimax value:

\begin{equation}
J_{\mathrm{LB}}(f,d)
\leq J_{\mathrm{L}}^\star(f,d)
\leq J_{\mathrm{L}}(\widehat{\varepsilon}_{\mathrm{c}}^\ast;f,d).
\label{eq:LB_relation}
\end{equation}

Equation~\eqref{eq:LB_relation} provides lower and upper bounds on the minimax value. Their difference bounds the maximum possible improvement through further optimization. A difference of at most 0.005~dB establishes the prescribed numerical accuracy for the specified model, reference set and discrete incidence angles. For an application-defined budget $\delta_{\mathrm{L}}$, a candidate with $J_{\mathrm{L}}(\widehat{\varepsilon}_{\mathrm{c}}^\ast;f,d) \leq\delta_{\mathrm{L}}$ establishes feasibility. Conversely, $J_{\mathrm{LB}}(f,d)>\delta_{\mathrm{L}}$ certifies infeasibility independently of the search algorithm. If the candidate objective exceeds the budget while the lower bound does not, neither conclusion follows from these two tests alone, and further refinement may be required. The critical depth is defined as the infimum of the observation depths at which the prescribed loss tolerance cannot be satisfied in Case A:

\begin{equation}
\begin{split}
d_{\mathrm{crit}}(f,\delta_{\mathrm{L}})
&=\inf\bigl\{d\in[0,d_{\mathrm{skin}}]:
\Omega_{\delta_{\mathrm{L}}}(f,d)=\varnothing\bigr\},
\end{split}
\label{eq:dcrit}
\end{equation}

where $d_{\mathrm{skin}}=1.3~\mathrm{mm}$ denotes the upper limit of the investigated observation depths specified in Section~\ref{subsec:evaluation_framework}. This definition identifies the onset of infeasibility without assuming monotonicity with depth, allowing the tolerance to become attainable again at greater depths. Different complex permittivities may be selected at different observation depths. If the tolerance can be satisfied throughout the investigated interval, the result is reported as $d_{\mathrm{crit}}>1.3~\mathrm{mm}$. An initial numerical sweep comprises 131 observation depths per frequency, ranging from 0 to 1.3~mm in increments of 0.01~mm. Starting from the surface, the candidate obtained at the preceding depth is evaluated at the next depth and compared with the corresponding analytical lower bound. A difference exceeding 0.005~dB initiates successive local grid refinements around this candidate. Whenever these refinements are insufficient to achieve the prescribed numerical accuracy, the full grid search described in Section~\ref{subsec:robust_specification} is performed. To establish feasibility between the sampled depths, each interval is subsequently examined by testing whether a common candidate satisfies the loss tolerance at both endpoints. For a fixed candidate and skin reference at a given frequency, incidence angle and polarization, the loss difference in Case A is described by Eq.~\eqref{eq:loss_decomposition}. Its interface contribution is independent of depth, while its propagation contribution varies linearly with depth because the normal attenuation coefficients are constant in homogeneous media. For fixed candidate and reference permittivities, the signed channel-loss difference is affine in depth. Its absolute value is therefore convex, so the maximum over a depth interval is attained at one of its endpoints. Taking the maximum over all selected reference models, incidence angles and polarizations preserves this property. For two observation depths $d_{\mathrm{a}}$ and $d_{\mathrm{b}}$, an intermediate depth is defined by $d_\lambda=(1-\lambda)d_{\mathrm{a}}+\lambda d_{\mathrm{b}}$, with $0\leq\lambda\leq1$. Convexity then yields

\begin{equation}
\begin{aligned}
J_{\mathrm{L}}(\widehat{\varepsilon}_{\mathrm{c}}^\ast;f,d_\lambda)
&\leq (1-\lambda)
J_{\mathrm{L}}(\widehat{\varepsilon}_{\mathrm{c}}^\ast;f,d_{\mathrm{a}})+\lambda
J_{\mathrm{L}}(\widehat{\varepsilon}_{\mathrm{c}}^\ast;f,d_{\mathrm{b}}).
\end{aligned}
\end{equation}

Consequently, a single candidate satisfying the prescribed budget at both interval endpoints also satisfies it throughout the intervening depth interval. This provides a continuous feasibility certificate for that candidate without requiring the minimax value obtained by optimizing separately at each depth to be convex or monotonic. Each candidate from a depth interval's endpoints is tested at both ends. If neither meets the tolerance at both ends, a midpoint candidate is computed and the interval subdivided as necessary. Adjacent verified segments establish continuous feasibility from the surface. To enclose the critical depth in the interval $[d_{\mathrm{crit}}^{-},d_{\mathrm{crit}}^{+}]$ (Eq.~\eqref{eq:dcrit}), feasibility must be verified continuously from the surface to the lower endpoint $d_{\mathrm{crit}}^{-}$. At the upper endpoint $d_{\mathrm{crit}}^{+}$, the analytical lower bound must exceed the tolerance,

\begin{equation}
J_{\mathrm{LB}}
(f,d_{\mathrm{crit}}^{+})
>
\delta_{\mathrm{L}}.
\end{equation}

Each finite critical-depth interval is refined to a width not exceeding $0.001~\mathrm{mm}$. At newly evaluated depths, the difference between the computed candidate objective and the analytical lower bound must not exceed $0.001$~dB. Surface infeasibility gives $d_{\mathrm{crit}}=0$. The critical-depth results for
tolerances of 1, 3 and 6~dB are based on these verified intervals. Their widths specify the numerical resolution of the reported critical depths.

\subsection{Comparison of material-selection strategies}
\label{subsec:benchmark_strategies}

To assess the benefit of robust inverse design at the specified depth, we compare the minimax target with direct selection of an individual skin reference model and a centroid with equal weighting of provenance groups. All strategies use the same frequency-specific reference set, incidence-angle grid, and $d=1.3~\mathrm{mm}$ endpoint.  Benchmark mismatches use Eq.~\eqref{eq:JL} separately for TE and TM by setting $\mathcal{P}=\{\mathrm{TE}\}$ or $\mathcal{P}=\{\mathrm{TM}\}$. The centroid is the same phantom candidate for both polarizations, whereas the best single-reference and robust targets are selected independently for each polarization. These benchmarks therefore differ from the joint TE/TM minimax optimization used for the primary material specification in Table~\ref{tab:headline}. For the single-reference benchmark, each reference permittivity is tested as the candidate in turn,

\begin{equation}
\varepsilon_{\mathrm{c}}^\ast=\varepsilon_m^\ast(f),
\qquad m\in\mathcal{M}(f).
\label{eq:single_model}
\end{equation}

The resulting minimum and maximum of $J_{\mathrm{L}}$ delimit the range of mismatches among the finite set of single-reference choices. A complementary benchmark, referred to as the centroid, accounts for the shared origin of several skin reference models by assigning equal weight to their provenance groups. Let $\mathcal{G}(f)$ denote the set of provenance groups represented in $\mathcal{M}(f)$, with $\mathcal{M}_g(f)$ containing the reference models assigned to group $g$ according to Supplementary Table~S4. Each reference model belongs to exactly one group. The cardinalities $|\mathcal{M}_g(f)|$ and $|\mathcal{G}(f)|$ denote the numbers of models within group $g$ and represented groups, respectively. The mean complex permittivity within each group is calculated as

\begin{equation}
\bar{\varepsilon}_{g}^{\ast}(f)
=
\frac{1}{|\mathcal{M}_g(f)|}
\sum_{m\in\mathcal{M}_g(f)}
\varepsilon_m^{\ast}(f).
\end{equation}

Averaging these group-specific values yields

\begin{equation}
\varepsilon_{\mathrm{cent,prov}}^{\ast}(f)
=
\frac{1}{|\mathcal{G}(f)|}
\sum_{g\in\mathcal{G}(f)}\bar{\varepsilon}_{g}^{\ast}(f).
\label{eq:centroid}
\end{equation}

Each provenance group thus contributes equally to the centroid, irrespective of the number of reference models it contains. This weighting defines a pointwise dielectric benchmark based on the documented source lineages. Since the Fresnel transmission and propagation attenuation depend nonlinearly on complex permittivity, averaging dielectric properties does not generally minimize the worst-case channel-loss mismatch. For the robust benchmark at depth $d$, the minimizing complex permittivities form the set

\begin{equation}
\mathcal{A}^\star(f,d)
=
\operatorname*{arg\,min}_{
\varepsilon_{\mathrm{c}}^\ast\in\mathcal{D}(f)
}
J_{\mathrm{L}}
(\varepsilon_{\mathrm{c}}^\ast;f,d),
\label{eq:robust_solution_set}
\end{equation}

For an attained infimum, the argmin set contains the minimizing permittivities. Numerical comparisons use a computed minimax candidate whose mismatch differs from the lower bound by no more than the reported numerical tolerance. Application-specific acceptance of this benchmark follows from membership in $\Omega_{\delta_{\mathrm{L}}}(f,d)$. A separate diagnostic, performed independently for TE and TM using the corresponding polarization-specific objective, examines whether phantom candidates with nearly identical worst-case agreement at the air--skin interface also provide comparable fidelity at finite depth. In the homogeneous forward model, setting $d=0$ removes the depth-dependent attenuation term and retains the interface-transmission response. Consequently, candidates with similarly small surface mismatch may still exhibit different normal attenuation coefficients, leading to different channel-loss responses at finite depths. The corresponding surface objective is

\begin{equation}
J_{\mathrm{L},\mathrm{surf}}(\varepsilon_{\mathrm{c}}^\ast;f)
=J_{\mathrm{L}}(\varepsilon_{\mathrm{c}}^\ast;f,0),
\label{eq:surface_objective}
\end{equation}

with best achievable value

\begin{equation}
J_{\mathrm{L},\mathrm{surf}}^\star(f)=J_{\mathrm{L}}^\star(f,0).
\label{eq:surface_optimum}
\end{equation}

A nonnegative allowance $\Delta_{\mathrm{surf}}$ above the exact surface infimum defines the near-optimal surface set

\begin{equation}
\begin{split}
\mathcal{N}_{\Delta_{\mathrm{surf}}}(f)
=\bigl\{&\varepsilon_{\mathrm{c}}^\ast\in\mathcal{D}(f):\\
&J_{\mathrm{L},\mathrm{surf}}(\varepsilon_{\mathrm{c}}^\ast;f)
\leq J_{\mathrm{L},\mathrm{surf}}^\star(f)+\Delta_{\mathrm{surf}}\bigr\}.
\end{split}
\label{eq:surface_set}
\end{equation}

To examine candidates with closely comparable agreement at the air--skin interface, the main analysis uses a surface allowance of $\Delta_{\mathrm{surf}}=0.05~\mathrm{dB}$. Additional values of 0.02 and 0.10~dB are considered to assess sensitivity to this selection threshold. The surface allowance specifies proximity to the best achievable surface mismatch, whereas the loss tolerance $\delta_{\mathrm{L}}$ specifies material acceptance. Numerical evaluation of $\mathcal{N}_{\Delta_{\mathrm{surf}}}(f)$ uses the computed surface optimum. Membership near the threshold is sensitive to the optimality gap of this computed value, so the sampled set approximates the theoretical set. Phantom candidates are sampled within the fixed plotting domain of Section~\ref{subsec:robust_specification} using a grid spacing of 0.15 along both the $\varepsilon_{\mathrm{c}}'$ and $\varepsilon_{\mathrm{c}}''$ axes. Candidates satisfying the numerical surface criterion are subsequently evaluated at $d=1.3~\mathrm{mm}$ while retaining their original complex permittivities. The resulting minimum and maximum define the mismatch range across the sampled grid. This comparison demonstrates that similar interface transmission can coexist with substantially different channel losses at finite depths.

\subsection*{Use of artificial intelligence}
ChatGPT (OpenAI) was used to improve the grammar, phrasing and readability of the manuscript and supplementary information. The authors retain full responsibility for the scientific content and final text.

\Needspace{13\baselineskip}
\section*{Funding}
The research presented in this paper was partially funded by the Deutsche Forschungsgemeinschaft, Project-ID 287022738, TRR 196 MARIE, for Projects M01, M03, S04, and S05. Additional support was provided by the Ministry of Culture and Science of the State of North Rhine-Westphalia (MKW NRW) through the terahertz.NRW project and by the German Federal Ministry of Research, Technology and Space (BMFTR) through the 6GEM research hub under grant number 16KISK039. The project ``terahertz.NRW'' receives funding from the program ``Netzwerke 2021,'' an initiative of the Ministry of Culture and Science of the State of North Rhine-Westphalia. The authors are solely responsible for the content of this publication. We also acknowledge financial support from the Open Access Publication Fund of the University of Duisburg-Essen.

\section*{Author contributions}
A.P. and A.A.A. jointly conceived the study, formulated the central research question, and contributed equally to its conceptual development. Both contributed to the analytical formulation, methodological development, and interpretation of the results. A.P. coordinated the numerical implementation and analysis and led the original manuscript draft. S.E. and Y.Z. contributed to computational analyses and the visualization of results, and provided scientific comments and revisions to the manuscript. A.P. and A.A.A. contributed to manuscript writing, review, and editing. M.J., J.M.J., D.E., and T.K. provided scientific guidance and supervision. All authors contributed ideas, discussed the results, reviewed the manuscript, and approved its final version.

\section*{Data and code availability}
The associated full-precision research dataset and implementation code are not publicly distributed with this preprint.

\section*{Competing interests}
The authors declare no competing interests.

\section*{Additional information}
Correspondence and requests for materials should be addressed to A.P. or A.A.A.

\begingroup
\setlength{\bibsep}{0.35em}
\raggedright
\emergencystretch=1em

\endgroup


\begin{thebibliography}{99}
\raggedright


\bibitem{Joseph2025} Joseph, L. \emph{et al.} Anthropomorphic torso phantoms and validation of their physical and digital twins for body-area radio-communication applications. \emph{IEEE Trans. Biomed. Eng.} \textbf{72,} 1858--1871 (2025). \url{https://doi.org/10.1109/TBME.2024.3524728}.

\bibitem{VanMulder2017} Van Mulder, T. J. S. \emph{et al.} High frequency ultrasound to assess skin thickness in healthy adults. \emph{Vaccine} \textbf{35,} 1810--1815 (2017). \url{https://doi.org/10.1016/j.vaccine.2016.07.039}.

\bibitem{Chahat2012} Chahat, N., Zhadobov, M. \& Sauleau, R. Broadband tissue-equivalent phantom for BAN applications at millimeter waves. \emph{IEEE Trans. Microw. Theory Techn.} \textbf{60,} 2259--2266 (2012). \url{https://doi.org/10.1109/TMTT.2012.2195196}.

\bibitem{GuraliucTMTT2014} Guraliuc, A. R., Zhadobov, M., De Sagazan, O. \& Sauleau, R. Solid phantom for body-centric propagation measurements at 60 GHz. \emph{IEEE Trans. Microw. Theory Techn.} \textbf{62,} 1373--1380 (2014). \url{https://doi.org/10.1109/TMTT.2014.2320691}.

\bibitem{GuraliucMobiHealth2014} Guraliuc, A. R., Zhadobov, M. \& Sauleau, R. PDMS-based skin-equivalent phantom for propagation studies in the 58--63 GHz range. In \emph{Proc. 4th Int. Conf. Wireless Mobile Communication and Healthcare (MobiHealth)} (2014). \url{https://doi.org/10.4108/icst.mobihealth.2014.257344}.

\bibitem{Khorshid2018} Khorshid, A. E., Alquaydheb, I. N., Eltawil, A. M. \& Kurdahi, F. J. Physical multi-layer phantoms for intra-body communications. \emph{IEEE Access} \textbf{6,} 42812--42821 (2018). \url{https://doi.org/10.1109/ACCESS.2018.2849709}.

\bibitem{Khorshid2019} Khorshid, A. E., Alquaydheb, I. N., Eltawil, A. M. \& Kurdahi, F. J. IBCFAP: Intra-Body Communications Five-Layers Arm Phantom Model. \emph{IEEE Access} \textbf{7,} 93701--93710 (2019). \url{https://doi.org/10.1109/ACCESS.2019.2925865}.

\bibitem{ProkschaIWMTS2023} Prokscha, A. \emph{et al.} A look through artificial human tissues at Ka-band and D-band. In \emph{Proc. 6th International Workshop on Mobile Terahertz Systems (IWMTS)}, 1--5 (IEEE, 2023). \url{https://doi.org/10.1109/IWMTS58186.2023.10207780}.

\bibitem{ZauberIRMMW2025} Zauber, T., Prokscha, A., Kolpatzeck, K., Kaiser, T. \& Czylwik, A. Scattering measurement for steep incident angles of skin phantoms at millimeter wave frequencies. In \emph{Proc. 50th International Conference on Infrared, Millimeter, and Terahertz Waves (IRMMW-THz)} (IEEE, 2025). \url{https://doi.org/10.1109/IRMMW-THz61557.2025.11319660}.

\bibitem{GhoshGupta2026} Ghosh, S. \& Gupta, B. Accurate equivalent human body phantom dielectric design for in vitro measurement of implantable antennas for biotelemetry. \emph{IEEE Trans. Dielectr. Electr. Insul.} \textbf{33,} 956--965 (2026). \url{https://doi.org/10.1109/TDEI.2025.3632236}.

\bibitem{GhoshEtAl2026} Ghosh, S., Ekpo, S. C., Elias, F., Alabi, S. \& Gupta, B. Estimation of complex permittivity of human body using transmission matrix method and FEM solver. \emph{Discov. Electron.} \textbf{3,} 59 (2026). \url{https://doi.org/10.1007/s44291-026-00213-y}.

\bibitem{Mizuno2024} Mizuno, M., Yamazaki, S. \& Nagaoka, T. Spectroscopic evaluation of epidermis-equivalent phantom in terahertz-frequency region. \emph{Radio Sci.} \textbf{59,} e2023RS007809 (2024). \url{https://doi.org/10.1029/2023RS007809}.

\bibitem{Yamazaki2025} Yamazaki, S. \emph{et al.} Investigation of reflectance, transmittance, and thermal response of skin and surface-textured skin-equivalent phantoms in the terahertz frequency range around 0.3 THz. \emph{Sci. Rep.} \textbf{15,} 45042 (2025). \url{https://doi.org/10.1038/s41598-025-28863-0}.

\bibitem{Prokscha2025HoneyBee} Prokscha, A. \emph{et al.} Perspectives on terahertz honey bee sensing. \emph{Sci. Rep.} \textbf{15,} 10638 (2025). \url{https://doi.org/10.1038/s41598-025-91630-8}.

\bibitem{Skos2026} Skos, K., Elayan, H., Elmaadawy, S., Kulakowski, P. \& Jornet, J. M. Nano-bio systems for intra-body sensing, actuation, and communication in the terahertz band and beyond. \emph{IEEE Commun. Surv. Tutor.} \textbf{28,} 6354--6383 (2026). \url{https://doi.org/10.1109/COMST.2026.3694302}.

\bibitem{IEEE8021562026} IEEE. \emph{IEEE Standard for Wireless Body Area Networks}, IEEE Std. 802.15.6-2026 (IEEE, 2026). https://standards.ieee.org/ieee/802.15.6/10941/.

\bibitem{Pellegrini2013} Pellegrini, A. \emph{et al.} Antennas and propagation for body-centric wireless communications at millimeter-wave frequencies: A review. \emph{IEEE Antennas Propag. Mag.} \textbf{55,} 262--287 (2013). \url{https://doi.org/10.1109/MAP.2013.6645205}.

\bibitem{Yang2015} Yang, K. \emph{et al.} Numerical analysis and characterization of THz propagation channel for body-centric nano-communications. \emph{IEEE Trans. THz Sci. Technol.} \textbf{5,} 419--426 (2015). \url{https://doi.org/10.1109/TTHZ.2015.2419823}.

\bibitem{Abbasi2016} Abbasi, Q. H. \emph{et al.} Terahertz channel characterization inside the human skin for nano-scale body-centric networks. \emph{IEEE Trans. THz Sci. Technol.} \textbf{6,} 427--434 (2016). \url{https://doi.org/10.1109/TTHZ.2016.2542213}.

\bibitem{Sasaki2022} Sasaki, K., Porter, E., Rashed, E. A., Farrugia, L. \& Schmid, G. Measurement and image-based estimation of dielectric properties of biological tissues---past, present, and future---. \emph{Phys. Med. Biol.} \textbf{67,} 14TR01 (2022). \url{https://doi.org/10.1088/1361-6560/ac7b64}.

\bibitem{Wang2021} Wang, J., Lindley-Hatcher, H., Chen, X. \& Pickwell-MacPherson, E. THz sensing of human skin: A review of skin modeling approaches. \emph{Sensors} \textbf{21,} 3624 (2021). \url{https://doi.org/10.3390/s21113624}.

\bibitem{Wu2015} Wu, T., Rappaport, T. S. \& Collins, C. M. Safe for generations to come: Considerations of safety for millimeter waves in wireless communications. \emph{IEEE Microw. Mag.} \textbf{16,} 65--84 (2015). \url{https://doi.org/10.1109/MMM.2014.2377587}.

\bibitem{FornesLeal2019} Fornes-Leal, A. \emph{et al.} Dielectric characterization of in vivo abdominal and thoracic tissues in the 0.5--26.5 GHz frequency band for wireless body area networks. \emph{IEEE Access} \textbf{7,} 31854--31864 (2019). \url{https://doi.org/10.1109/ACCESS.2019.2903481}.

\bibitem{Sipus2021} \v{S}ipu\v{s}, Z., \v{S}u\v{s}njara, A., Skrivervik, A. K., Poljak, D. \& Bosiljevac, M. Influence of uncertainty of body permittivity on achievable radiation efficiency of implantable antennas---stochastic analysis. \emph{IEEE Trans. Antennas Propag.} \textbf{69,} 6894--6905 (2021). \url{https://doi.org/10.1109/TAP.2021.3070594}.

\bibitem{Ghosh2025Link} Ghosh, S., Ekpo, S. C., Elias, F., Alabi, S. \& Gupta, B. Effects of dielectric properties of human body on communication link margins and specific absorption rate of implanted antenna system. \emph{Sensors} \textbf{25,} 3498 (2025). \url{https://doi.org/10.3390/s25113498}.

\bibitem{Pickwell2004} Pickwell, E., Cole, B. E., Fitzgerald, A. J., Pepper, M. \& Wallace, V. P. In vivo study of human skin using pulsed terahertz radiation. \emph{Phys. Med. Biol.} \textbf{49,} 1595--1607 (2004). \url{https://doi.org/10.1088/0031-9155/49/9/001}.

\bibitem{Pickwell2005} Pickwell, E. \emph{et al.} Simulating the response of terahertz radiation to basal cell carcinoma using ex vivo spectroscopy measurements. \emph{J. Biomed. Opt.} \textbf{10,} 064021 (2005). \url{https://doi.org/10.1117/1.2137667}.

\bibitem{Chan2018} Chan, K. Y. \& Ramer, R. Novel concept of detecting basal cell carcinoma in skin tissue using a continuous-wave millimeter-wave rectangular glass filled probe. \emph{Med. Devices: Evidence Res.} \textbf{11,} 275--285 (2018). \url{https://doi.org/10.2147/MDER.S168338}.

\bibitem{Zaytsev2015APL} Zaytsev, K. I., Kudrin, K. G., Karasik, V. E., Reshetov, I. V. \& Yurchenko, S. O. In vivo terahertz spectroscopy of pigmentary skin nevi: Pilot study of non-invasive early diagnosis of dysplasia. \emph{Appl. Phys. Lett.} \textbf{106,} 053702 (2015). \url{https://doi.org/10.1063/1.4907350}.

\bibitem{Sasaki2017} Sasaki, K., Mizuno, M., Wake, K. \& Watanabe, S. Monte Carlo simulations of skin exposure to electromagnetic field from 10 GHz to 1 THz. \emph{Phys. Med. Biol.} \textbf{62,} 6993--7010 (2017). \url{https://doi.org/10.1088/1361-6560/aa81fc}.

\bibitem{Saxena2024} Saxena, S. \emph{et al.} Limitations of effective medium models for tissue phantoms in the THz frequency range. \emph{Sci. Rep.} \textbf{14,} 22968 (2024). \url{https://doi.org/10.1038/s41598-024-70590-5}.

\bibitem{Truong2013} Truong, B. C. Q., Tuan, H. D., Kha, H. H. \& Nguyen, H. T. Debye parameter extraction for characterizing interaction of terahertz radiation with human skin tissue. \emph{IEEE Trans. Biomed. Eng.} \textbf{60,} 1528--1537 (2013). \url{https://doi.org/10.1109/TBME.2013.2237908}.

\bibitem{Truong2015Skin} Truong, B. C. Q., Tuan, H. D., Wallace, V. P., Fitzgerald, A. J. \& Nguyen, H. T. The potential of the double Debye parameters to discriminate between basal cell carcinoma and normal skin. \emph{IEEE Trans. THz Sci. Technol.} \textbf{5,} 990--998 (2015). \url{https://doi.org/10.1109/TTHZ.2015.2485208}.

\bibitem{Multifaceted} Prokscha, A. \emph{et al.} Multifaceted Free-Space Tissue Sensing up to 100 GHz. \emph{IEEE Access} \textbf{12,} 107670--107688 (2024). \url{https://doi.org/10.1109/ACCESS.2024.3439605}.

\bibitem{Gabriel1996} Gabriel, S., Lau, R. W. \& Gabriel, C. The dielectric properties of biological tissues: III. Parametric models for the dielectric spectrum of tissues. \emph{Phys. Med. Biol.} \textbf{41,} 2271--2293 (1996). \url{https://doi.org/10.1088/0031-9155/41/11/003}.

\bibitem{Alekseev2007} Alekseev, S. I. \& Ziskin, M. C. Human skin permittivity determined by millimeter wave reflection measurements. \emph{Bioelectromagnetics} \textbf{28,} 331--339 (2007). \url{https://doi.org/10.1002/bem.20308}.

\bibitem{Chahat2011} Chahat, N., Zhadobov, M., Augustine, R. \& Sauleau, R. Human skin permittivity models for millimetre-wave range. \emph{Electron. Lett.} \textbf{47,} 427--428 (2011). \url{https://doi.org/10.1049/el.2011.0349}.

\bibitem{Zhekov2019} Zhekov, S. S., Franek, O. \& Pedersen, G. F. Dielectric properties of human hand tissue for handheld devices testing. \emph{IEEE Access} \textbf{7,} 61949--61959 (2019). \url{https://doi.org/10.1109/ACCESS.2019.2914863}.

\bibitem{Ghodgaonkar2000} Ghodgaonkar, D. K., Gandhi, O. P. \& Iskander, M. F. Complex permittivity of human skin in vivo in the frequency band 26.5--60 GHz. In \emph{Proc. IEEE Antennas Propag. Soc. Int. Symp.}, Vol.~2, 1100--1103 (2000). \url{https://doi.org/10.1109/APS.2000.875414}.

\bibitem{Mirbeik2018} Mirbeik-Sabzevari, A., Ashinoff, R. \& Tavassolian, N. Ultra-wideband millimeter-wave dielectric characteristics of freshly excised normal and malignant human skin tissues. \emph{IEEE Trans. Biomed. Eng.} \textbf{65,} 1320--1329 (2018). \url{https://doi.org/10.1109/TBME.2017.2749371}.

\bibitem{Alabaster2005} Alabaster, C. Permittivity measurements on human skin in the millimetre wave band. In \emph{Proc. 18th Int. Conf. Applied Electromagnetics and Communications (ICECom)} (2005). \url{https://doi.org/10.1109/ICECOM.2005.205042}.

\bibitem{Hwang2003} Hwang, H., Yim, J., Cho, J.-W., Cheon, C. \& Kwon, Y. 110 GHz broadband measurement of permittivity on human epidermis using 1 mm coaxial probe. In \emph{IEEE MTT-S International Microwave Symposium Digest} \textbf{1,} 399--402 (2003). \url{https://doi.org/10.1109/MWSYM.2003.1210961}.

\bibitem{TruongThesis2015} Truong, B. C. Q. \emph{Dielectric Modelling of Human Skin and Breast Tissue in Terahertz Frequencies: Potential Application to Cancer Detection}. Ph.D. dissertation, University of Technology Sydney (2015), Table~4.1. \url{https://opus.lib.uts.edu.au/bitstream/10453/43388/2/02whole.pdf}.

\bibitem{Wallace2006} Wallace, V. P. \emph{et al.} Terahertz pulsed spectroscopy of human basal cell carcinoma. \emph{Appl. Spectrosc.} \textbf{60,} 1127--1133 (2006). \url{https://doi.org/10.1366/000370206778664635}.

\bibitem{Echchgadda2013} Echchgadda, I. \emph{et al.} Using a portable terahertz spectrometer to measure the optical properties of in vivo human skin. \emph{J. Biomed. Opt.} \textbf{18,} 120503 (2013). \url{https://doi.org/10.1117/1.JBO.18.12.120503}.

\bibitem{Zaytsev2015} Zaytsev, K. I., Gavdush, A. A., Chernomyrdin, N. V. \& Yurchenko, S. O. Highly accurate in vivo terahertz spectroscopy of healthy skin: Variation of refractive index and absorption coefficient along the human body. \emph{IEEE Trans. THz Sci. Technol.} \textbf{5,} 817--827 (2015). \url{https://doi.org/10.1109/TTHZ.2015.2460677}.

\bibitem{Mattana2024} Mattana, E., Lodi, M. B., Simone, M., Mazzarella, G. \& Fanti, A. Cole--Cole model for the dielectric characterization of healthy skin and basal cell carcinoma at THz frequencies. \emph{IEEE Open J. Eng. Med. Biol.} \textbf{5,} 600--606 (2024). \url{https://doi.org/10.1109/OJEMB.2024.3438562}.

\bibitem{Christ2025} Christ, A. \emph{et al.} Human skin model from 15 GHz to 110 GHz. \emph{Bioelectromagnetics} \textbf{46,} e70025 (2025). \url{https://doi.org/10.1002/bem.70025}.

\bibitem{BalanisAEM2012} Balanis, C. A. \emph{Advanced Engineering Electromagnetics}, 2nd edn (Wiley, Hoboken, NJ, 2012).

\bibitem{Ashworth2009} Ashworth, P. C. \emph{et al.} Terahertz pulsed spectroscopy of freshly excised human breast cancer. \emph{Opt. Express} \textbf{17,} 12444--12454 (2009). \url{https://doi.org/10.1364/OE.17.012444}.

\bibitem{Wilmink2011} Wilmink, G. J. \emph{et al.} Development of a compact terahertz time-domain spectrometer for the measurement of the optical properties of biological tissues. \emph{J. Biomed. Opt.} \textbf{16,} 047006 (2011). \url{https://doi.org/10.1117/1.3570648}.

\bibitem{Piro2016} Piro, G. \emph{et al.} Terahertz electromagnetic field propagation in human tissues: A study on communication capabilities. \emph{Nano Commun. Netw.} \textbf{10,} 51--59 (2016). \url{https://doi.org/10.1016/j.nancom.2016.07.010}.

\bibitem{YehLayeredMedia} Yeh, P. \emph{Optical Waves in Layered Media} (Wiley, New York, 1988).

\bibitem{Costanzo2021} Costanzo, S., Cioffi, V., Qureshi, A. M. \& Borgia, A. Gel-like human mimicking phantoms: Realization procedure, dielectric characterization and experimental validations on microwave wearable body sensors. \emph{Biosensors} \textbf{11,} 111 (2021). \url{https://doi.org/10.3390/bios11040111}.

\bibitem{Mirbeik2019} Mirbeik-Sabzevari, A. \& Tavassolian, N. Ultrawideband, stable normal and cancer skin tissue phantoms for millimeter-wave skin cancer imaging. \emph{IEEE Trans. Biomed. Eng.} \textbf{66,} 176--186 (2019). \url{https://doi.org/10.1109/TBME.2018.2828311}.

\bibitem{Lacik2016} L\'a\v{c}\'ik, J., Hebelka, V., V\'elim, J., Raida, Z. \& Puskely, J. Wideband skin-equivalent phantom for V- and W-band. \emph{IEEE Antennas Wirel. Propag. Lett.} \textbf{15,} 211--213 (2016). \url{https://doi.org/10.1109/LAWP.2015.2438441}.
\end{thebibliography}
\end{document}